\documentclass[10pt, aps, groupedaddress, twocolumn, nofootinbib, prd, tightenlines, floatfix, superscriptaddress]{revtex4-1}
\usepackage{latexsym}
\usepackage{amsmath}
\usepackage{amsfonts}
\usepackage{amsbsy}
\usepackage{amssymb}
\usepackage{epsfig}
\usepackage{mathrsfs}
\usepackage{epsfig}
\usepackage{bbm}
\usepackage{color, verbatim, graphicx, tikz}
\usepackage[colorlinks=true, citecolor=cyan, linkcolor=.,urlcolor=.,]{hyperref}

\def\eqref#1{(\textcolor{blue}{\ref{#1}})}

\begin{document}

\title{{\bf Hopf link volume simplicity constraints in spin foam models}}

\author{Mehdi Assanioussi}
\email[]{mehdi.assanioussi@desy.de}
\email[]{mehdi.assanioussi@fuw.edu.pl}
\affiliation{II. Institute for Theoretical Physics, University of Hamburg\\ Luruper Chaussee 149, 22761 Hamburg, Germany}
\affiliation{Faculty of Physics, University of Warsaw, Pasteura 5, 02-093 Warsaw, Poland.}
\author{Benjamin Bahr}
\email[]{benjamin.bahr@desy.de}
\affiliation{II. Institute for Theoretical Physics, University of Hamburg\\ Luruper Chaussee 149, 22761 Hamburg, Germany}

\date{\today}

\begin{abstract}
In this article we consider specific bivector geometries which arise in the large-spin limit of the extension  of the Engle-Pereira-Rovelli-Livine spin foam model for quantum gravity by Kaminski, Kisielowski and Lewandowski. We address the implementation of volume simplicity constraints, which are required to ensure that a $4d$ metric can be reconstructed from the bivector geometry. We find that the necessary conditions are closely related, but not quite equal to the Hopf link volume simplicity constraints introduced in earlier works. We estimate the number of independent geometricity conditions for arbitrary bivector geometries, and find that they always agree with the number of Hopf links on the graph minus one, suggesting that the geometricity conditions can generically be formulated by deformation of the Hopf link volume simplicity constraints. 
\end{abstract}


\maketitle

\section{Motivation}

Spin Foam Models are a tentative proposal for a discrete path-integral expression of quantum gravity \cite{Baez:2002aw}. They rest on the on-shell-equivalence of the Einstein-Hilbert action with the Holst  action \cite{Holst:1995pc}, which in turn can be written as a topological theory of a $\text{Spin}(1,3)$-connection for Lorentzian signature, or a $\text{Spin}(4)$-connection for Riemannian signature, and a bivector-valued 2-form $B$ with the BF action \cite{Horowitz:1989ng}, plus a set of simplicity constraints
\begin{eqnarray}\label{Eq:Simplicity_Continuum}
B_{[\mu\nu]}^{[IJ]}\;\sim\;\epsilon^{IJ}{}_{KL}e^{K}_{\mu}e^{L]}_{\nu},
\end{eqnarray}

\noindent which allow the recovery of a vierbein $e^I_\mu$ from the $B$-field, and hence a $4d$ metric $g_{\mu\nu}=\eta_{IJ} e^I_\mu,e^J_\nu$.  

Spin Foam models then proceed with a discretization of the continuum theory on a lattice, which takes the form of either an embedded or an abstract 2-complex. Early versions employed  the 2-skeleton of the dual to a $4d$ triangulation \cite{Barrett:1997gw,Engle:2007wy,Freidel:2007py}, while it was later generalised to general 2-complexes \cite{Kaminski:2009fm}. 

The quantization of the discretized data yields a formally topological state sum, on which the simplicity constraints (\ref{Eq:Simplicity_Continuum}) are implemented \emph{a posteriori}. In the model by Engle et al \cite{Engle:2007wy}, defined on a $4d$ triangulation $\mathcal{T}$, the simplicity constraints take the form of a set of conditions on bivectors associated to the 2-cells of the 2-complex, or, equivalently, to the triangles of $\mathcal{T}$.  The simplicity constraint conditions can be shown to ensure that these bivectors are those of a  geometric $4d$ simplex, hence assigning a Regge-like geometry to the triangulation. In that sense, the simplicity constraints ensure that the discretized $B$-field is coming from a $4d$ metric, allowing the connection to classical (discrete) General Relativity \cite{Barrett:2009gg}\cite{Conrady:2008mk}. 

For more general 2-complexes, the situation is less clear. While one can always reconstruct $3d$ polytopes from the bivector data \cite{Bianchi:2010gc}, it has been observed that for 2-complexes dual to a more general cellular decomposition of space-time, the simplicity constraints are insufficient to constrain the 2-complexes data enough so that a reconstruction of a geometric polytope is possible \cite{Bahr:2015gxa, Bahr:2017ajs, Dona:2017dvf, SimonePaper}. Indeed, there might not exist a $4d$ metric, and even in the limit of large quantum numbers, the excess degrees of freedom survive, which suggests that the classical limit of the theory contains more fields than original GR. Some effort has been made to relate these excess degrees of freedom to other geometric quantities, e.g.~components of the torsion tensor \cite{Bahr:2015gxa}, and in a general analysis it was found that these degrees of freedom generically manifest themselves as shape-mismatch of polytope areas \cite{Dona:2017dvf}.

In some examples it has been shown that the excess degrees of freedom are a consequence of an insufficient implementation of the so-called volume-simplicity constraints, which comprise a subset of the whole set of constraints \cite{Perez:2012wv}. These are not required in the case of a 4-simplex geometry, which is why there is no problem to leave them out in the original models. However, for more general polytopes, they have to be implemented, which so far no Spin foam model provides. In fact, for general polytopes, their form is not known.  Therefore, it would be desirable to have a generic form of sufficient implementation of the full set of simplicity constraints at hand, to ensure the reduction of spin foam data to $4d$ geometries.

For a set of hypercuboidal boundaries, a form of these constraints has been identified \cite{Bahr:2017ajs}. They can be formulated as an equivalence of certain quantities, so-called ``Hopf link volumes'', associated to Hopf links in boundary graphs around the vertices in the 2-complex. While these constraints ensure the reconstruction of $4d$ polytopes from bivector data, their form for general bivector geometries on arbitrary graphs has not been proven.

Indeed, in the following article, we show that the Hopf link volume simplicity constraints as formulated in \cite{Bahr:2017ajs} are \emph{not quite} equivalent to the conditions of geometricity. However, on the graph we are considering, one can show that their number always agree, and it appears that the conditions of geometricity are deformed versions of the Hopf link volume simplicity constraints. This suggests that geometricity might still be related to the Hopf links in boundary graphs, although the precise relation is still open at this time.\\

In this article, we proceed as follows: firstly, we recap the basic notions of spin network functions and bivector geometries in section \ref{Sec:SF_Basics}. Then, in section \ref{Sec:Anvil_Geometries}, we consider a very specific set of data on a hypercubic graph, which we dubbed ``anvil geometries''. These are generalizations of the hypercuboidal and frustal geometries employed in earlier works, in particular those in renormalization computations \cite{Bahr:2016hwc, Bahr:2017klw, Bahr:2018gwf}. These geometries are naturally of interest as generalizations of these works. We then carefully compute the resulting bivector geometries arising at the stationary and critical points of the amplitudes. These are the geometries which are expected to play a dominant role in the classical limit of the theory, since they are the ones not exponentially suppressed in the region of large quantum numbers. We further show that not all of these bivector geometries are geometric, i.e.~allow for a reconstruction of a $4d$ polytope, and we identify the geometricity conditions for the reconstruction. We further estimate the number of geometricity conditions on arbitrary graphs and arbitrary spin network data. After that, in section \ref{Sec:HLVolumes}, we compare the geometricity conditions with the Hopf link volume simplicity constraints proposed in \cite{Bahr:2017ajs}, and we find that they only agree when certain additional conditions are satisfied, i.e.~only for a subset of anvil geometries. However, we show via a perturbative analysis of the two sets of constraints that for data close to the hypercuboidal configurations, the Hopf link volume simplicity constraints can be a good approximation of the geometricity conditions. We conclude with a summary and some outlook in section \ref{Sec:Summary}. 

\section{Spin Foam basics}\label{Sec:SF_Basics}

Our investigations rest on findings in the Riemannian signature Spin Foam Model by Engle, Pereira, Livine and Rovelli (EPRL) \cite{Engle:2007wy}, however there is a host of models on the market for either signature, and many of the following statements also apply to those \cite{Barrett:1997gw, Freidel:2007py, CubulationSpinFoamThiemann2008, Baratin:2011hp}. The EPRL model was extended by Kaminski, Kisielowski and Lewandowski to arbitrary 2-complexes, resulting in the EPRL-KKL model \cite{Kaminski:2009fm}, which is what we will be working with in this article. For a general recap of the Spin Foam formalism see \cite{Perez:2012wv} and references therein, and see \cite{Bahr:2012qj} for the holonomy formulation, in which these models can be cast.

Since the continuum form of the simplicity constraints are local, their discrete version is local to the 2-complex, i.e.~are conditions on the variables in the vicinity around a vertex $v$, i.e.~0-cell. This data can be represented on a graph, which arises as the intersection of an open 3-ball around $v$ with the 2-complex. This is also called the boundary graph around $v$. The data on a graph $\Gamma$ consists of spins and intertwiners, corresponding to a generalization of Penrose's spin networks, which serve as the boundary data for the theory, connecting Spin Foams to Loop Quantum Gravity on the kinematic level. The space of intertwiners admits an overcomplete basis given by the Livine-Speziale coherent states, the data of which are $3d$ normals associated to ordered pairs of nodes for which there is a link in $\Gamma$.

Concretely, consider a graph $\Gamma$ (which can be thought of as embedded in $S^3$) with nodes $n\in N(\Gamma)$ and oriented links $\ell\in L(\Gamma)$. The data for a LS-coherent state is given by
\begin{eqnarray}\label{Eq:SpinNetworkData}
\Big\{j_{ab},\,\vec{n}_{ab}\Big\}_{(ab)=\ell\in L(\Gamma)}
\end{eqnarray}

\noindent where $j_{ab}=j_{ba}\in \frac{1}{2}\mathbb{N}$ are half-integers labelling a spin associated to the link, and $\vec{n}_{ab}\in S^2$ labels normal vectors satisfying the closure constraints
\begin{eqnarray}\label{Eq:Closure_3D}
\sum_{b: (ab)\in L(\Gamma)}[a,b] j_{ab}\vec{n}_{ab}\;=\; 0\qquad\text{for all }a\in N(\Gamma).
\end{eqnarray}

\noindent The sum ranges over all nodes $b$ which are connected to $a$ by a link, and $[a,b]=-[b,a]$ is equal to $1$ if the link orientation is going from $a$ to $b$. By Minkowski's theorem, data satisfying (\ref{Eq:Closure_3D}) corresponds to a unique (up to translation, barring degenerate configurations) $3d$ convex polytope with areas $j_{ab}$ and normals $ \vec{n}_{ab}$. This allows one to interpret the spin network data, which is summed over in the representation picture of spin foam models, as sum over discrete geometries, where the boundary of a space-time region is a collection of connected $3d$ polytopes which are internally flat \cite{Bianchi:2010gc}.

\subsection{From spin network data to bivectors}

The main part of spin foam models are their amplitudes, which are local functions of the data (\ref{Eq:SpinNetworkData}) associated to the cells of the complex. These specify the probabilities of transitions between different piecewise-flat $3d$ geometries. The vertex amplitude $\mathcal{A}_v$ associated to 0-cells (vertices) $v$ is, in the EPRL-KKL model, given by an integral over group elements $g_a^\pm\in SU(2)$. The other amplitudes are straightforward functions of spin network data, which is why the spin foam path integral can be viewed as summing over configurations of spins, intertwiners and group elements.

For each configuration, a set of bivectors can be constructed from the group elements and the spin network data via
\begin{eqnarray}
B_{ab}\;\sim\;j_{ab}\big(g_a^+\triangleright\vec{n}_{ab},\,g_a^-\triangleright\vec{n}_{ab}\big)
\end{eqnarray}

\noindent where the isomorphism $\mathbb{R}^3\oplus\mathbb{R}^3\simeq \mathbb{R}^4\wedge\mathbb{R}^4$ has been used\footnote{See appendix \ref{App:SelfDualConventions}.}, as well as the canonical action $\triangleright$ of $SU(2)$ on $\mathbb{R}^3$. Note that, in general, $B_{ab}\neq -B_{ba}$, which means that there is no unique bivector associated to the (oriented)  links of boundary graphs, which correspond to 2-cells in the complex. However, for large spins, it can be shown that the EPRL-KKL-amplitude is exponentially suppressed for all but those configurations, which satisfy $B_{ab}= -B_{ba}$, or, equivalently:
\begin{eqnarray}\label{Eq:CritStat_SU2}
g_{a}^\pm\triangleright \vec{n}_{ab}\;=\;-g_b^\pm\triangleright \vec{n}_{ba}.
\end{eqnarray}

\noindent These equations arise in the extended stationary phase, which is why the configurations satisfying these conditions are called critical and stationary. They allow to construct a discrete version of the bivectors from the spin foam data.

\subsection{From bivectors to geometricity}

In the EPRL model, the boundary graphs $\Gamma=K_5$ around vertices are always complete graphs in five vertices. Disregarding degenerate data, the critical and stationary bivector data allows to reconstruct a 4-simplex at each vertex, i.e.~there exist a geometric 4-simplex whose 2-face-bivectors are exactly the $B_{ab}$. Also, the amplitude at these configurations can be related to the exponential of the Regge action \cite{Barrett:2009gg}\cite{Magliaro:2011dz}\cite{Han:2017xwo} for many configurations, which raises the hope that the classical limit of spin foam models is related to discrete general relativity.

In the EPRL-KKL model, the graph $\Gamma$ is arbitrary, e.g.~those which are dual boundary graphs of more general polytopes. For these, the equations (\ref{Eq:CritStat_SU2}) do not necessarily allow the reconstruction of a 4-dimensional polytope. Hence the model is not constrained enough to restrict the bivector geometries to metrics (see for instance \cite{Asante:2020qpa}). For specific sets of bivectors and graphs, this can be traced back to the missing implementation of a discrete version of the volume simplicity constraints \cite{Bahr:2017ajs}, and we suspect this to be the case in general. 

In general, the failure of existence of a geometric 4-polytope can be traced back to shape-mismatch of 2-faces: The constraints on bivectors in the EPRL-KKL models allow to construct a $3d$ polytope which is embedded in $\mathbb{R}^4$ for each node of the graph $\Gamma$, and for each pair of nodes $a, b$ connected by an oriented link $(ab)$ the corresponding $3d$ polytope are sharing a common 2-face. The geometry of that 2-face is a $2d$ polygon which can be constructed using the bivectors around either node. The crucial point is that this polygon is only constrained to have the same area, $4d$ orientation, and inner dihedral angles for either reconstruction \cite{Dona:2017dvf}. For polygons more complicated than triangles, this does not completely fix the geometry. Ensuring the complete match of the face does then imply the existence of a $4d$ polytope, which does not have to be convex, and can have self-intersections, see e.g.~\cite{Bahr:2018vvq} (we ignore degenerate configurations, which can e.g.~result in zero-volume polytopes). \\

The open question is then: what are the constraints needed to be imposed on the $B_{ab}$, such that the reconstruction of a $4d$ polytope can be ensured? We essentially have the following situation:
\begin{eqnarray*}
\begin{array}{cl}
j_{ab},\vec{n}_{ab} & \quad \text{spin network data}\\[15pt]
\downarrow & \quad\text{critical \& stationary conditions}\\[15pt]
B_{ab} & \quad \text{bivector geometry}\\[15pt]
\downarrow & \text{geometricity conditions}\\[15pt]
P & \quad \text{ $4d$ geometric polytope}
\end{array}
\end{eqnarray*}

\noindent The exact form of the geometricity conditions is the main unsolved issue which this article is concerned with. 

\subsection{Hopf link volume simplicity and geometricity}

Geometricity, i.e.~complete matching of $2d$ faces, can be achieved by demanding that the edge lengths of $2d$ polygons, constructed from either $3d$ polyhedron, agree. As equality of lengths, these constraints are straightforward but overcomplete, and the computation of lengths from bivectors is non-trivial. Also, edge lengths are part of the $4d$ polytope, and not part of the boundary graph of the polytope. In particular, for arbitrary graphs, a polytope might not even exist. Hence, specifying lengths requires to postulate the existence of a dual polytope before stating the condition for its existence. 

\begin{figure}[hbt!]
\begin{center}
\includegraphics[scale=0.35]{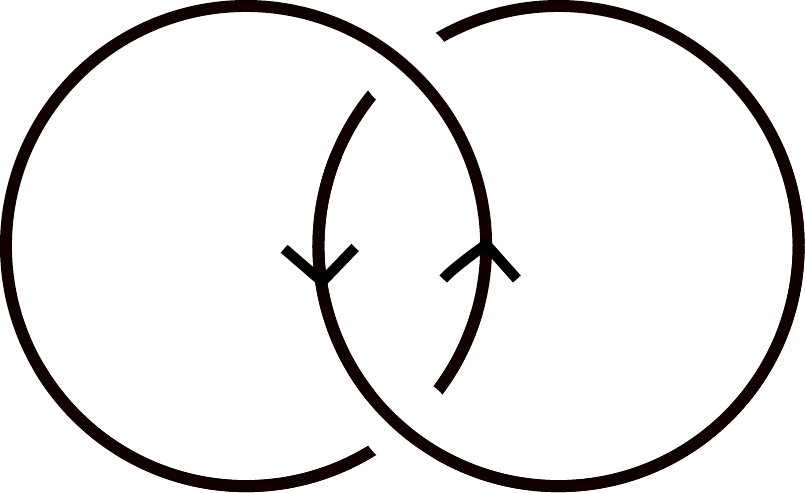}
\caption{A Hopf link, consisting of two closed loops with relative winding number 1.}\label{Fig:HopfLink}
\end{center}
\end{figure}

Therefore, it would be desirable to have a form of the constraints which ensure geometricity, directly in terms of the bivectors and other quantities only inherent to the boundary graph. These could then be quantized directly and implemented on the quantum level. Also, this would be desirable from the point of view of having a connection to the original simplicity constraints, which in the continuum are given by
\begin{eqnarray}
\epsilon_{IJKL}B^{IJ}_{\mu\nu}B^{KL}_{\nu\rho}\;=\;V\epsilon_{\mu\nu\sigma\rho},
\end{eqnarray}

\noindent where $V$ is the $4$-volume computed from the vierbeins in (\ref{Eq:Simplicity_Continuum}). For special geometries on a particular graph, these constraints have been explicitly constructed in \cite{Bahr:2017ajs}, and they can be phrased as follows: project the boundary graph on a plane and consider the \emph{Hopf links} $H$, i.e.~closed loops of links in the projected graph which have relative winding number 1 (see figure \ref{Fig:HopfLink}). For these, compute the \emph{Hopf link volume}
\begin{eqnarray}
V_H\;:=\;
\frac{1}{6}\sum_{C\subset H} \epsilon_{IJKL}B_{(ab)}^{IJ}B_{(cd)}^{KL} 
\end{eqnarray}

\noindent where the sum ranges over all crossings $C$ in the Hopf link, which correspond to where links $(ab)$ and $(cd)$ cross. \footnote{Here the orientations of links $(ab)$, $(cd)$ have to be chosen such that the crossing has crossing number 1, see \cite{Bahr:2017ajs} for details.} The condition of  equality of all Hopf link volumes $V_H$ for all Hopf links $H$ in the graph is called \emph{Hopf link volume simplicity constraints} (HLVS constraints):
\begin{eqnarray}
V_{H}\;=\;V_{H'}\qquad\text{for all }H,H'.
\end{eqnarray}

\noindent For the examples in \cite{Bahr:2017ajs}, it could be shown that geometricity and HLVS are equivalent, providing an elegant criterion for geometricity directly from the data on the boundary graph. Furthermore, the HLVS condition can be formulated on any other graph as well.

Still, it was so far unclear if geometricity and HLVS constraints are equivalent for all bivector geometries on all graphs, and in fact, in the following we present an example where these two sets of conditions are almost equivalent, but not quite.

\section{Anvil geometries}\label{Sec:Anvil_Geometries}

We are interested in the transition of a 3d geometry of a cuboid $Q_1$ at the initial time step, to a differently-sized cuboid $Q_2$ at the finite time step. This is what we will call a 4d anvil for now. 

A 4d geometry which would produce such a transition is the join between the two cuboids $Q_{1,2}$, placed in parallel 3d hyperplanes in $\mathbb{R}^4$ separated by a distance $h$, which can be considered to be the 4-dimensional height. The 4d polytope in question is then given by the join
\begin{align}\label{Eq:Definition4dPolytopeJoin}
P
\;=\;
\big\{
\,p=\lambda q_1+(1-\lambda)q_2\,
\big|
\,q_i\in Q_i,\lambda\in[0,1]\,
\big\}.
\end{align}

\begin{figure}[hbt!]
\begin{center}
\includegraphics[scale=0.35]{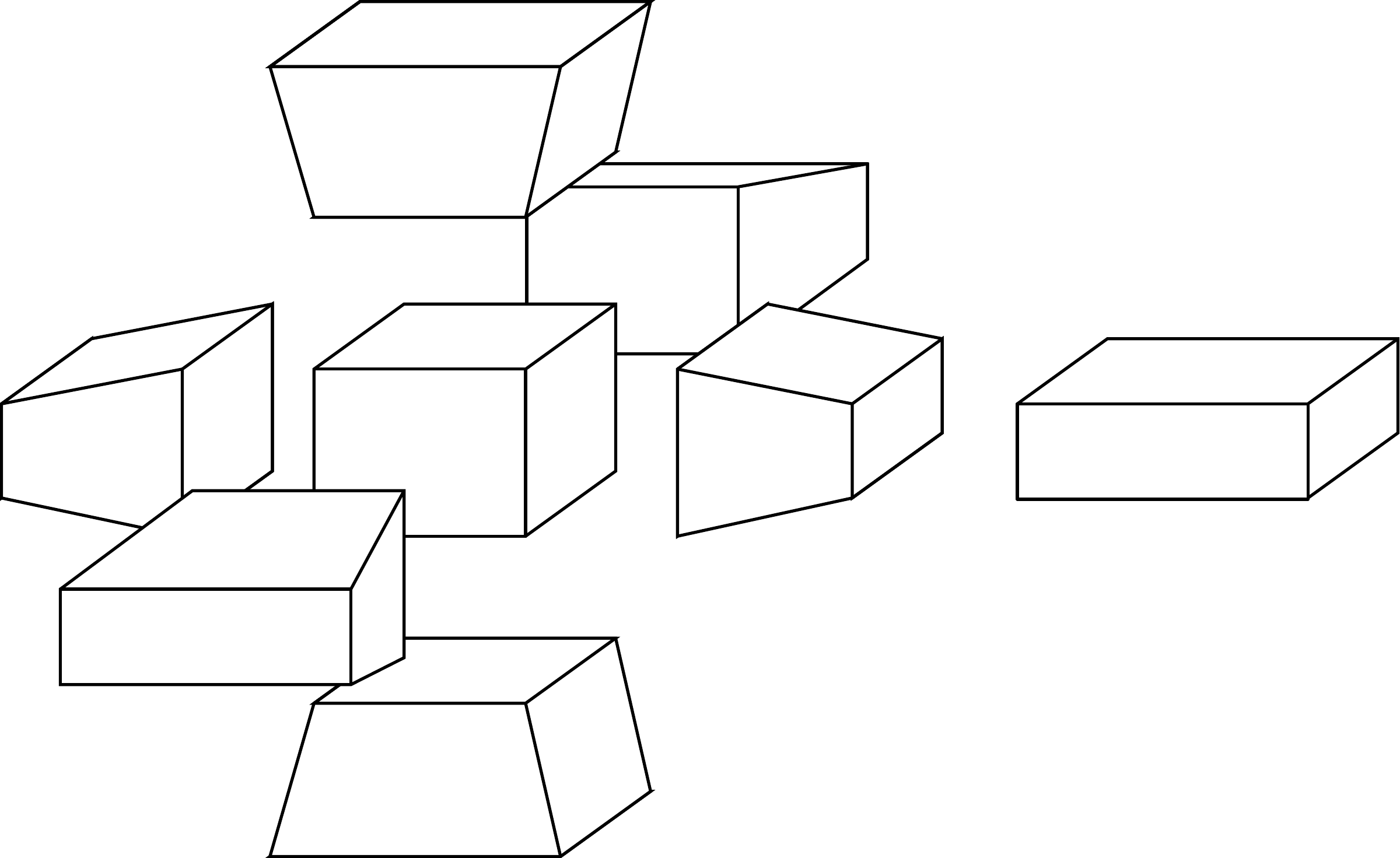}
\caption{The boundary of a $4d$ anvil, folded up into $\mathbb{R}^3$. The cuboids correspond to $Q_1$, $Q_2$; while the $3d$ anvils are $T_1,\ldots T_6$.}\label{Fig:Anvil_4d_Geometric}
\end{center}
\end{figure}

\begin{figure}[hbt!]
\begin{center}
\includegraphics[scale=0.65]{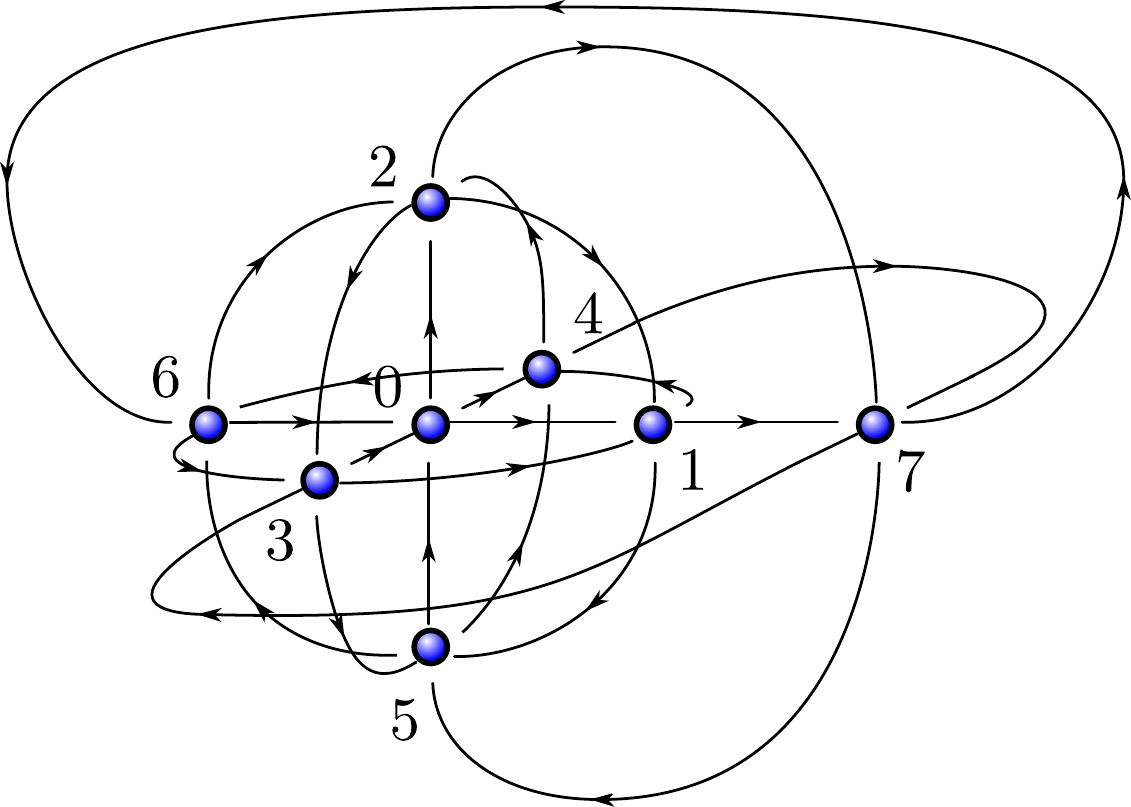}
\caption{The boundary graph of a $4d$ anvil, folded up into $\mathbb{R}^3$. Every node corresponds to a $3d$ polytope in figure \ref{Fig:Anvil_4d_Geometric}. There is a link whenever two of those are glued together in the boundary of $P$.}\label{Fig:Anvil_4d_Graph}
\end{center}
\end{figure}

\begin{figure}[hbt!]
\begin{center}
\includegraphics[scale=0.45]{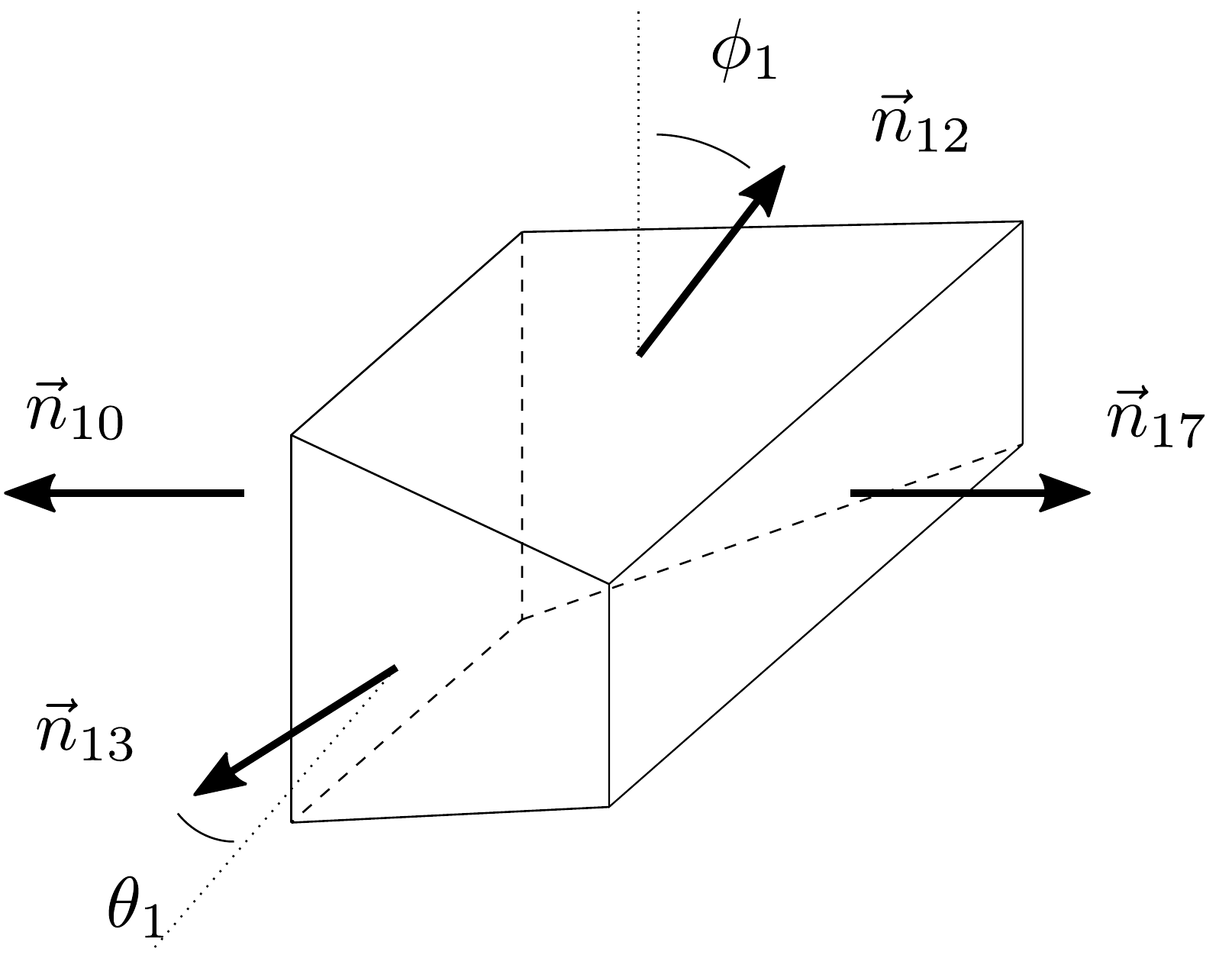}
\caption{Pre-bivector geometry of the node $1$ in figure \ref{Fig:Anvil_4d_Graph}.}\label{Fig:Figure_3dAnvil}
\end{center}
\end{figure}

\noindent Since each of the two cubes is completely determined by three parameters, one needs seven parameters to describe a 4d anvil, i.e.~six lengths and $h$. The boundary of $P$, unfolded into 3d space, is depicted in figure \ref{Fig:Anvil_4d_Geometric}. The boundary graph $\Gamma$ consists of eight nodes, depicted in figure \ref{Fig:Anvil_4d_Graph}. Two of them correspond to $Q_1$ and $Q_2$, while the remaining six $T_i$, $i=1,\ldots,6$ correspond to ``time-like'' polytopes resembling 3d anvils, depicted in figure \ref{Fig:Figure_3dAnvil}. Due to symmetry, one has 
\begin{align}\label{Eq:SymmetryOfPolytopes}
T_i=T_{7-i}
\end{align}
\noindent  within the boundary of the anvil.

Conversely, we prescribe the ``pre-bivector''-3d boundary data in terms of areas $j_{ab}=j_{ba}$ and normals $\vec{n}_{ab}$, where $a$, $b$ are the nodes of $\Gamma$, connected by a link $(ab)$. A cuboid (up to rotation) has completely fixed normals, which only leaves three areas, which are equivalent to specifying the three edge lengths of the cuboids. 

A 3d anvil as in figure \ref{Fig:Figure_3dAnvil} is specified by 5 parameters, which can be seen as follows: the node has six links leaving it, i.e.~there are six faces, where due to symmetry, however, two pairs are equal, respectively, i.e.~$j_{12}=j_{15}$ and $j_{13}=j_{14}$. The six normals are given by
\begin{align*}
\vec{n}_{10}
&=
\left(
\begin{array}{c}
-1\\0\\0
\end{array}
\right),\,
\vec{n}_{12}
=
\left(
\begin{array}{c}
\sin\phi_1\\ \cos\phi_1\\0
\end{array}
\right),\,
\vec{n}_{15}
=
\left(
\begin{array}{c}
\sin\phi_1\\-\cos\phi_1\\0
\end{array}
\right),
\\[5pt]
\vec{n}_{17}
&=
\left(
\begin{array}{c}
1\\0\\0
\end{array}
\right),\,
\vec{n}_{13}
=
\left(
\begin{array}{c}
-\sin\theta_1\\0\\ \cos\theta_1
\end{array}
\right),\,
\vec{n}_{14}
=
\left(
\begin{array}{c}
-\sin\theta_1\\0\\ -\cos\theta_1
\end{array}
\right).
\end{align*}

\noindent The closure condition (\ref{Eq:Closure_3D}) translates to
\begin{align}\label{Eq:Closure_3dAnvil}
j_{17}-j_{01}-2j_{13}\sin\theta_1+2j_{12}\sin\phi_1=0
\end{align}

\noindent This allows, in principle, to express one of the spins or one of the angles in terms of the other variables. So there are five variables in total.

Due to the symmetry (\ref{Eq:SymmetryOfPolytopes}), there are only three different 3d anvils, and the specification of these completely determines unambiguously the areas of the two cuboids $Q_{1,2}$. Since one has, e.g.~$j_{12}=j_{21}$, one has three additional constraints, fixing spins of one 3d anvil in terms of those of another. In total, there are hence 12 free parameters to specify a 4d-anvil pre-bivector geometry. These can be parameterised by six spins and six angles 
\begin{align}\label{Eq:ParametersPreBivectorGeometry}
j_1,\ldots, j_6,\;\phi_i,\,\theta_i,\quad i=1,2,3
\end{align}
\noindent  the following way:
\begin{align}\label{Eq:Spins}
\begin{aligned}
j_1
\;&:=\;
\frac{j_{01}+j_{17}}{2}
\;=\;
\frac{j_{06}+j_{67}}{2}
\\[5pt]
j_2
\;&:=\;
\frac{j_{02}+j_{27}}{2}
\;=\;
\frac{j_{05}+j_{57}}{2}\\[5pt]
j_3
\;&:=\;
\frac{j_{03}+j_{37}}{2}
\;=\;
\frac{j_{04}+j_{47}}{2}\\[5pt]
j_4
\;&:=\;
j_{12}=j_{26}=j_{56}=j_{15}\\[5pt]
j_5
\;&:=\;
j_{13}=j_{36}=j_{46}=j_{14}\\[5pt]
j_6
\;&:=\;
j_{23}=j_{35}=j_{45}=j_{24},
\end{aligned}
\end{align}

\noindent The original spins $j_{ab}$ can be obtained with the help of the  three \emph{intermediate spins}
\begin{align}\label{ls}
r_1
\;&:=\;
j_5\sin\theta_1
\,-\,
j_4\sin\phi_1\\[5pt]
r_2
\;&:=\;
j_4\sin\theta_2
\,-\,
j_6\sin\phi_2\\[5pt]
r_3
\;&:=\;
j_6\sin\theta_3
\,-\,
j_5\sin\phi_3
\end{align}

\noindent with which one has e.g.
\begin{align*}
j_{01}
\;=\;
j_1-r_1,
\qquad 
j_{17}
\;=\;
j_1+r_1,
\end{align*}
\noindent and so on. Note that the variables (\ref{Eq:ParametersPreBivectorGeometry}) already parameterize gauge-invariant data, i.e.~the $j_{ab}$ computed from them automatically satisfy the closure condition, e.g.~(\ref{Eq:Closure_3dAnvil}).

We can arrange the spin network data such that we group it to each of the three intermediate $3d$ anvils, i.e.
\begin{eqnarray}\label{Eq:GICS_Data}
\begin{aligned}
\text{anvil 1: }&j_1,\,j_4,\,j_5,\,\phi_1,\,\theta_1\\[5pt]
\text{anvil 2: }&j_2,\,j_4,\,j_6,\,\phi_2,\,\theta_2\\[5pt]
\text{anvil 3: }&j_3,\,j_5,\,j_6,\,\phi_3,\,\theta_3
\end{aligned}
\end{eqnarray}

\noindent These are the variables which one needs in order to describe three $3d$-anvils, such that their respective areas match according to the combinatorics of the graph \ref{Fig:Anvil_4d_Graph}. Alternatively, each of the $3d$ anvils can be described by 5 metric variables (two rectangles and a $3d$ height) and three area matching constraints:

\subsection{Metric variables}

\noindent Each of the three sets of  gauge-invariant spin foam data variables (\ref{Eq:GICS_Data}) describes, due to Minkowski's theorem, a $3d$ anvil, which is the join of two aligned rectangles (see figure \ref{Fig:Figure_3dAnvil}), which can also be described by four side lengths and the 3d height.

\begin{figure}[hbt!]
\begin{center}
\includegraphics[scale=0.45]{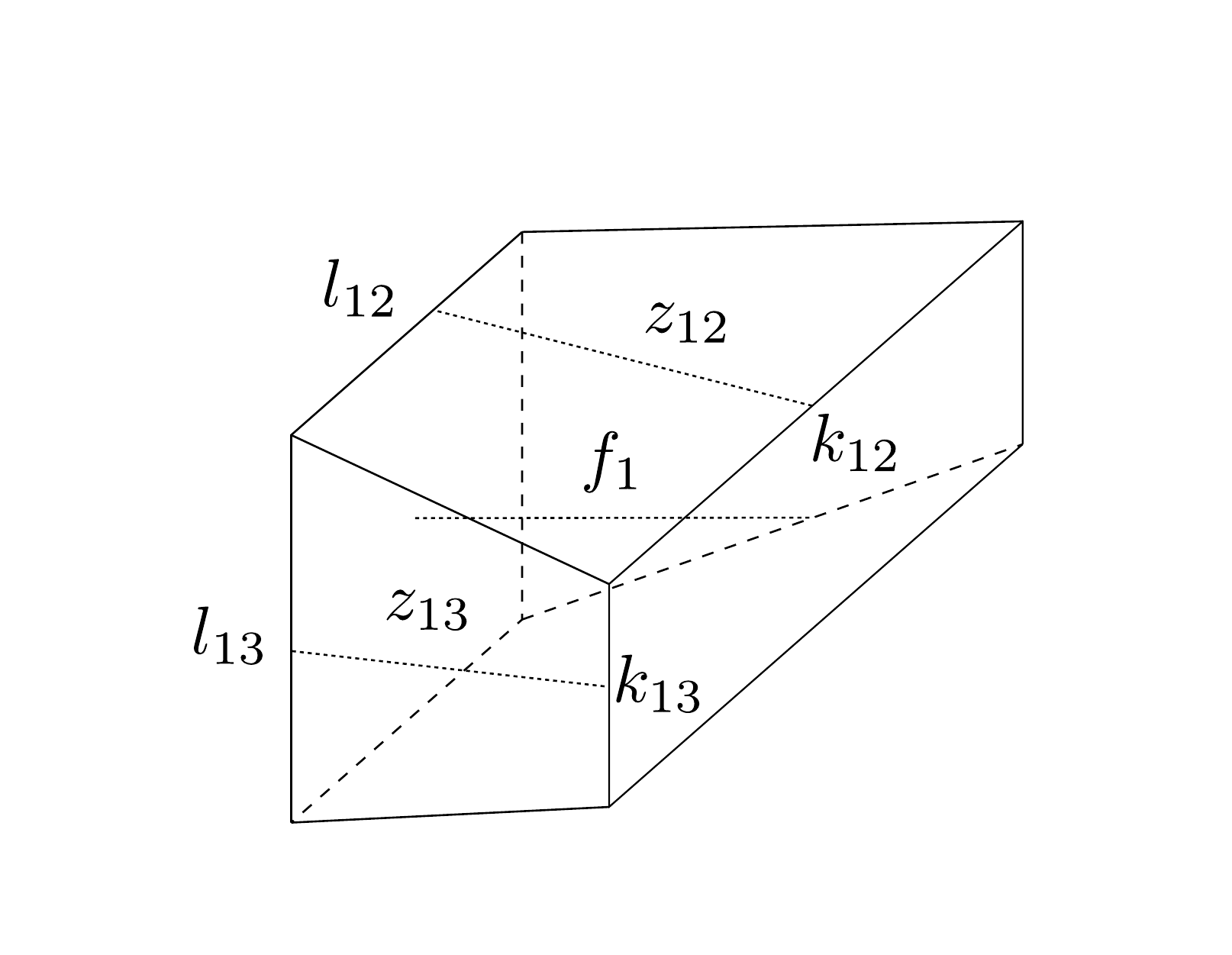}
\caption{Metric variables of the anvil associated to node $1$ in figure \ref{Fig:Anvil_4d_Graph}.}\label{Fig:Figure_3dAnvil_metric}
\end{center}
\end{figure}

The notation is such that the lengths $k_{ij}$ and $l_{ij}$ are, respectively, the top and bottom lengths of the rectangles in $3d$ frustum $i$, which touch the $3d$ frustum $j$. Also, $f_i$ denotes the $3d$ height of anvil $i$. So the metric variables are
\begin{eqnarray}\label{Eq:MetricVariables}
k_{ij},k_{ik},l_{ij},l_{ik},f_i,\qquad \{i,j,k\}\;=\;\{1,2,3\}.
\end{eqnarray}

\noindent E.g.~for anvil 1, these are $k_{12}$, $k_{13}$, $l_{12}$, $l_{13}$, and $f_1$. These are related to the spin foam data (\ref{Eq:GICS_Data}) via:
\begin{eqnarray}
\begin{aligned}
&j_1\;=\;\frac{k_{12}k_{13}+l_{12}l_{13}}{2},&\\[5pt]
&j_4\;=\;\frac{k_{12}+l_{12}}{2}z_{12},\quad j_5\;=\;\frac{k_{13}+l_{13}}{2}z_{13},&\\[5pt]
&\cos\phi_1\;=\;\frac{f_1}{z_{12}},\quad\sin\phi_1\;=\;\frac{l_{13}-k_{13}}{2z_{12}},&\\[5pt]
&\cos\theta_1\;=\;\frac{f_1}{z_{13}},\quad\sin\theta_1\;=\;\frac{k_{12}-l_{12}}{2z_{13}}&\\[5pt]
\end{aligned}
\end{eqnarray}
\noindent with
\begin{eqnarray}
\begin{aligned}
z_{12}^2\;&=\;f_1^2\,+\,\left(\frac{l_{13}-k_{13}}{2}\right)^2,\\[5pt]
z_{13}^2\;&=\;f_1^2\,+\,\left(\frac{k_{12}-l_{12}}{2}\right)^2.
\end{aligned}
\end{eqnarray}

\noindent Here the variables $z_{ij}$ are the $2d$ heights of the trapezoid faces in the $3d$ anvil $i$ touching the anvil $j$. 

The variables and relations for the other two anvils are given by a cyclic permutation of the indices $1\to 2\to 3\to 1$, and simultaneous $6\to 5\to 4\to 6$. This gives 15 variables in total. However, since the areas of gluing faces between anvils are given by spins, and therefore constrained to agree seen from either anvil, we have the three constraints

\begin{eqnarray}\label{Eq:MetricVariableConstraints}
\begin{aligned}
j_4\;&=\;\frac{k_{12}+l_{12}}{2}z_{12}\;=\;\frac{k_{21}+l_{21}}{2}z_{21}\\[5pt]
j_5\;&=\;\frac{k_{13}+l_{13}}{2}z_{13}\;=\;\frac{k_{31}+l_{31}}{2}z_{31}\\[5pt]
j_6\;&=\;\frac{k_{23}+l_{23}}{2}z_{23}\;=\;\frac{k_{32}+l_{32}}{2}z_{32}
\end{aligned}
\end{eqnarray}

\noindent bringing the number of independent variables down to 12.

\subsection{Bivector geometries}

\noindent The 12 variables (\ref{Eq:GICS_Data}), or equivalently (\ref{Eq:MetricVariables}) with (\ref{Eq:MetricVariableConstraints}), describe a set of geometries larger than 4d anvils, which are given by 7 parameters. In what follows we describe the five constraints required so that the data (\ref{Eq:GICS_Data}) describe a $4d$ anvil. 

One condition is that the spin network data actually describes a bivector geometry. This is equivalent to requiring that the critical and stationary equations of the asymptotic analysis of the vertex amplitude to be satisfied \cite{Barrett:2009gg}. These conditions can be rephrased as requiring the existence of $SU(2)$-elements $g_a$ which satisfy (\ref{Eq:CritStat_SU2}), and can be expressed in terms of the $j_n$, $n=1,\ldots, 6$, and $\phi_i$, $\theta_i$, $i=1,2,3$. Firstly, we observe that the corresponding rotations $R_a\in SO(3)$ have to be gauge-fixed, which we choose to do via
\begin{align*}
R_0
\;=\;
\mathbbm{1}_{3\times 3}.
\end{align*}

\noindent From this, in order to satisfy (\ref{Eq:CritStat_SU2}) for $a=0$, $b=1,2,3$, one can read off that $R_i$, $i=1,2,3$ have to be of the form
\begin{align}\label{Eq.AnsatzRotationMatrix123}
\begin{aligned}
R_1
\;=\;
R_{\vec{e}_1,\chi_1}
\;=\;
\left(
\begin{array}{ccc}
1 & 0 & 0 \\
0 & \cos\chi_1 & -\sin\chi 1 \\
0 & \sin\chi_1 & \cos\chi_1
\end{array}
\right)\\[5pt]
R_2
\;=\;
R_{\vec{e}_2,\chi_2}
\;=\;
\left(
\begin{array}{ccc}
\cos\chi_2 & 0 & \sin\chi_2 \\
0 & 1 & 0 \\
-\sin\chi_2 & 0 & \cos\chi_2
\end{array}
\right)\\[5pt]
R_3
\;=\;
R_{\vec{e}_3,\chi_3}
\;=\;
\left(
\begin{array}{ccc}
\cos\chi_3 & -\sin\chi_3 & 0 \\
\sin\chi_3 & \cos\chi_3 & 0 \\
0 & 0 & 1
\end{array}
\right)
\end{aligned}
\end{align}

\noindent Furthermore, we have, due to symmetry,
\begin{align*}
R_{7-i}
\;=\;
\big(
R_i
\big)^{-1}\qquad i=1,2,3,
\end{align*}

\noindent as well as $R_7=R_0$. From this, we get a set of three equations for each of the pairs $(ab)=(12)$, $(13)$, $(23)$. These equations are:
\begin{align}\label{Eq:CSEquations}
\begin{aligned}
\sin\phi_1
\;&=\;
-\cos\chi_2\,\cos\theta_2\\[5pt]
\cos\chi_1\,\cos\phi_1
\;&=\;
\sin\theta_2\\[5pt]
\sin\chi_1\,\cos\phi_1
\;&=\;
\sin\chi_2\,\cos\theta_2\\[5pt]
\sin\phi_2
\;&=\;
-\cos\chi_3\,\cos\theta_3\\[5pt]
\cos\chi_2\,\cos\phi_2
\;&=\;
\sin\theta_3\\[5pt]
\sin\chi_2\,\cos\phi_2
\;&=\;
\sin\chi_3\,\cos\theta_3\\[5pt]
\sin\phi_3
\;&=\;
-\cos\chi_1\,\cos\theta_1\\[5pt]
\cos\chi_3\,\cos\phi_3
\;&=\;
\sin\theta_1\\[5pt]
\sin\chi_3\,\cos\phi_3
\;&=\;
\sin\chi_1\,\cos\theta_1
\end{aligned}
\end{align}

\noindent Note that in the first, second, and third triple of equations, only two are independent, since they are effectively equations on $S^2$ because the vectors $\vec{n}_{ab}$ are normalised. We then get six equations in total after introducing three new angles $\chi_i$, and they constitute constraint equations on the $\phi_i$, $\theta_i$:
\begin{align}\label{Eq:EquationsAndConstraints}
\begin{aligned}
\cos\chi_1
\;&=\;
\frac{\sin\theta_2}{\cos\phi_1}
\;=\;
-\frac{\sin\phi_3}{\cos\theta_1}\\[5pt]
\cos\chi_2
\;&=\;
\frac{\sin\theta_3}{\cos\phi_2}
\;=\;
-\frac{\sin\phi_1}{\cos\theta_2}\\[5pt]
\cos\chi_3
\;&=\;
\frac{\sin\theta_1}{\cos\phi_3}
\;=\;
-\frac{\sin\phi_2}{\cos\theta_3}
\end{aligned}
\end{align}

\noindent One note on the signs here: from the definition one can see that $\phi_i,\theta_i\in(-\frac{\pi}{2},\frac{\pi}{2})$, hence all cosines are positive. However, sines can have either sign, coinciding with the sign of the angle. Therefore, from (\ref{Eq:EquationsAndConstraints}) we can immediately read off the conditions
\begin{eqnarray*}
\text{sgn}(\theta_2)\;&=&\;-\text{sgn}(\phi_3)\\[5pt]
\text{sgn}(\theta_3)\;&=&\;-\text{sgn}(\phi_1)\\[5pt]
\text{sgn}(\theta_1)\;&=&\;-\text{sgn}(\phi_2)
\end{eqnarray*}

\noindent Furthermore, we have conditions on $\phi_i$, $\theta_i$ in terms of inequalities coming from the fact that cosines have norms not larger than 1. If all of these conditions are satisfied, and knowing that the relative signs of the $\chi_i$ are fixed by (\ref{Eq:CSEquations}) to agree since all cosines are positive, we can use (\ref{Eq:EquationsAndConstraints}) to determine $\cos\chi_i$, with $\chi_i\in(-\pi,\pi)$. This uniquely determines all $\chi_i$ up to a common sign flip, which demonstrates that we have two solutions (or one solution with multiplicity $2$ in the case where all $\chi_i=0$, which however corresponds to a degenerate configuration). This pair of solutions $\pm\chi_i$ constitutes the two solutions determining the tentative $4d$ polytope, such that the $4d$ dihedral angles at the faces are given by $\chi_i$ \cite{Dona:2017dvf}.

In terms of the metric variables (\ref{Eq:MetricVariables}), the additional constraints can be understood in terms of the matching of $2d$ dihedral angles. First we note that with (\ref{Eq:CSEquations}) we have
\begin{eqnarray}
\begin{aligned}
\frac{\sin\chi_1}{\sin\chi_2}\;=\;\frac{\cos\theta_2}{\cos\phi_1}
\end{aligned}
\end{eqnarray}
\noindent and cyclic permutations, from which we can infer that
\begin{eqnarray}\label{Eq:CSRelationsCosines}
\prod_{i=1}^3\frac{\cos\phi_i}{\cos\theta_i}\;=\;\frac{z_{13}}{z_{31}}\,\frac{z_{32}}{z_{23}}\,\frac{z_{21}}{z_{12}}\;=\;1.
\end{eqnarray}
\noindent Also, from (\ref{Eq:EquationsAndConstraints}) we get
\begin{eqnarray}
\sin\phi_3\frac{\cos\theta_1}{\cos\phi_1}\;=\;-\sin\theta_2
\end{eqnarray}
\noindent and cyclic permutations, from which we get
\begin{eqnarray}\label{Eq:CSRelationsSines}
\prod_{i=1}^3\sin\phi_i\;=\;-\prod_{i=1}^3\sin\theta_i.
\end{eqnarray}

\begin{figure}[hbt!]
\begin{center}
\includegraphics[scale=0.35]{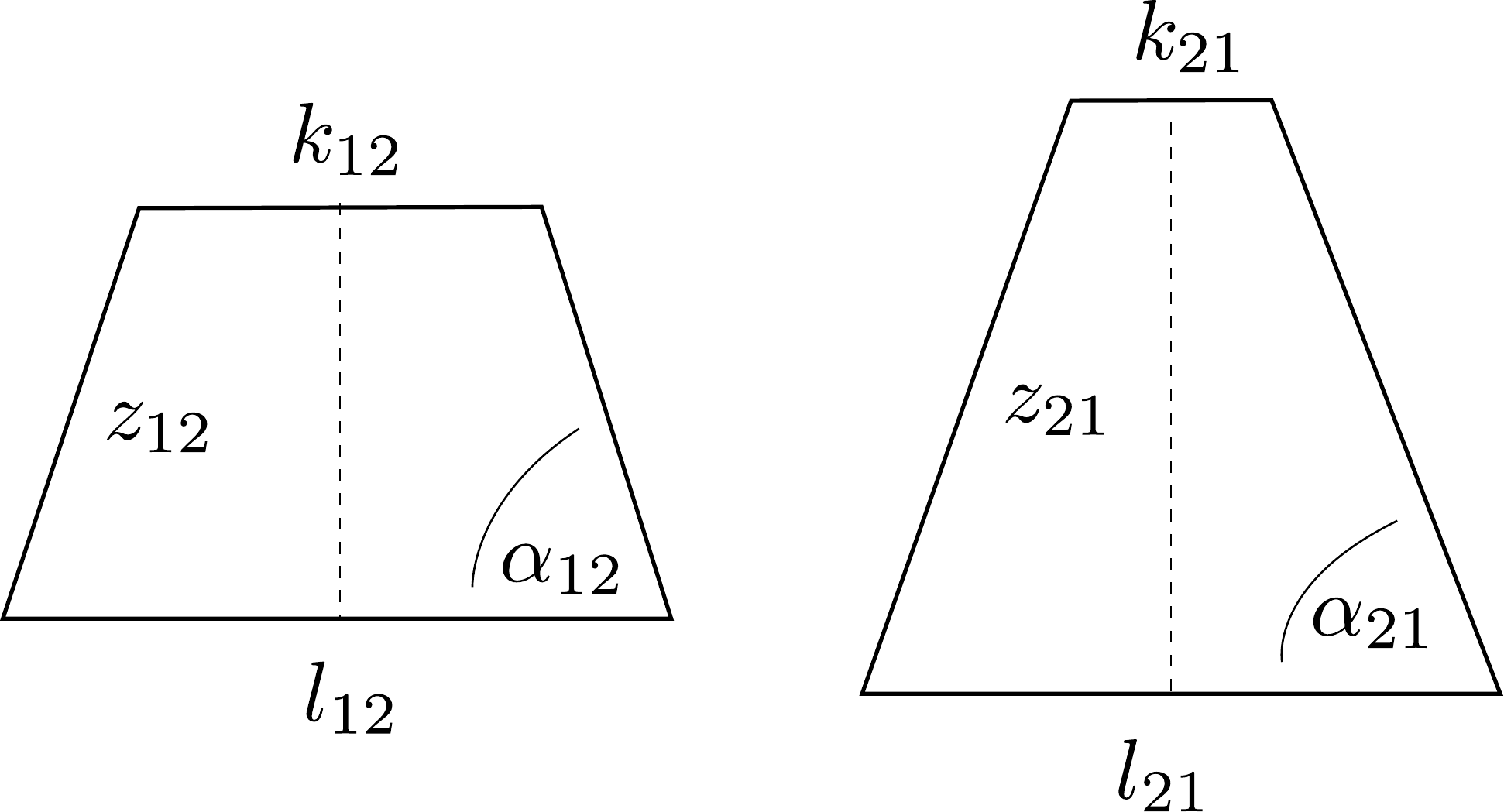}
\caption{Two matching trapezoid faces on node 1 and 2, respectively. The areas coincide, the angles $\alpha_{12}$, $\alpha_{21}$ coincide if (\ref{Eq:CSEquations}) are satisfied. Total shape matching is given when $z_{12}=z_{21}$, which is geometricity.}\label{Fig:Two_Trapezoids}
\end{center}
\end{figure}

\noindent Next, we consider the trapezoid face between $3d$ anvil 1 and 2 (figure \ref{Fig:Two_Trapezoids}). The interior $2d$ dihedral angles are $\alpha_{12}$ and $\alpha_{21}$. We get
\begin{eqnarray}
\begin{aligned}
\cot\alpha_{12}\;&=\;\frac{l_{12}-k_{12}}{2z_{12}}\;=\;\sin\theta_1\frac{\cos\phi_1}{\cos\theta_1}\\[5pt]
\cot\alpha_{21}\;&=\;\frac{l_{21}-k_{21}}{2z_{12}}\;=\;-\sin\phi_2\frac{\cos\theta_2}{\cos\phi_2}.
\end{aligned}
\end{eqnarray}

\noindent With (\ref{Eq:EquationsAndConstraints}) and (\ref{Eq:CSRelationsCosines}) we can see that
\begin{eqnarray}
\cot\alpha_{12}\;=\cot\alpha_{21},
\end{eqnarray}

\noindent and since the cotangent is monotonous on $(0,\pi)$, which is the range of possible values for the $\alpha_{ij}$, we get that, due to symmetry:
\begin{eqnarray}\label{Eq:InnerDihedral2dAnglesAgree}
\alpha_{ij}\;=\;\alpha_{ji}\qquad\text{for }i\neq j\in\{1,2,3\}.
\end{eqnarray}

\noindent In other words, the $2d$ inner dihedral angles of opposing faces agree for all faces. This is in line with the findings of \cite{Dona:2017dvf} for twisted spike geometries. 

\subsection{Geometricity constraints \& reconstruction}

The 9-dimensional set of bivector geometries, described by the 12 variables (\ref{Eq:GICS_Data}) and three constraints (\ref{Eq:EquationsAndConstraints}), is still more than the 7-dimensional set of $4d$ anvil geometries. In particular, we need two more constraint equations.  We will describe these on the bivector data, i.e.~we assume that (\ref{Eq:EquationsAndConstraints}) hold.

The geometricity constraints can be interpreted as shape-matching conditions of the trapezoid faces of the touching $3d$ anvils. These are determined by three numbers, e.g.~the total area $j_i$, one $2d$ dihedral angle $\alpha_{ij}$, and the height $z_{ij}$. The areas agreeing is a consequence of the spin foam variables. The angles $\alpha_{ij}=\alpha_{ji}$ agreeing is a consequence of the critical and stationary equations (\ref{Eq:CSEquations}). Therefore, geometricity in terms of face matching is a consequence of equality of the remaining heights of glued faces, i.e.
\begin{eqnarray}\label{Eq:GeometricityConditions}
z_{ij}\;=\;z_{ji}\qquad\text{for }i\neq j\in\{1,2,3\}.
\end{eqnarray}

\noindent Note that these are three equations, of which only two are independent, due to (\ref{Eq:CSRelationsCosines}). Two more conditions bring 9 down to 7, which describes the set of geometric $4d$ anvils.

We first define three new variables $h_i$, which are well-defined on the 9-dimensional set of bivector geometries:
\begin{eqnarray}\label{Eq:Definition4dHeight}
h_i\;:=\;f_i\sin\chi_i\;,\quad i\in\{1,2,3\}.
\end{eqnarray}

\noindent We also introduce the shorthand notation
\begin{eqnarray}\label{Eq:Definition_X_i}
\begin{aligned}
X_1\;&:=\;\cos\phi_2\sin\chi_2\;=\;\cos\theta_3\sin\chi_3,\\[5pt]
X_2\;&:=\;\cos\phi_3\sin\chi_3\;=\;\cos\theta_1\sin\chi_1,\\[5pt]
X_3\;&:=\;\cos\phi_1\sin\chi_1\;=\;\cos\theta_2\sin\chi_2.
\end{aligned}
\end{eqnarray}

\noindent Using (\ref{Eq:Definition4dHeight}) and (\ref{Eq:MetricVariables}), one can show that
\begin{eqnarray}
\begin{aligned}\label{Eq:PropertiesOfHeights}
h_1\;&=\;z_{12}X_3\;=\;z_{13}X_2\\[5pt]
h_2\;&=\;z_{23}X_1\;=\;z_{21}X_3\\[5pt]
h_3\;&=\;z_{31}X_2\;=\;z_{32}X_1
\end{aligned}
\end{eqnarray}

\noindent This shows that the geometricity constraints can be equivalently described as
\begin{eqnarray}\label{Eq:GeometricityConstraintAlternative}
h_1\;=\;h_2\;=\;h_3.
\end{eqnarray}

\begin{figure}[hbt!]
\begin{center}
\includegraphics[scale=0.55]{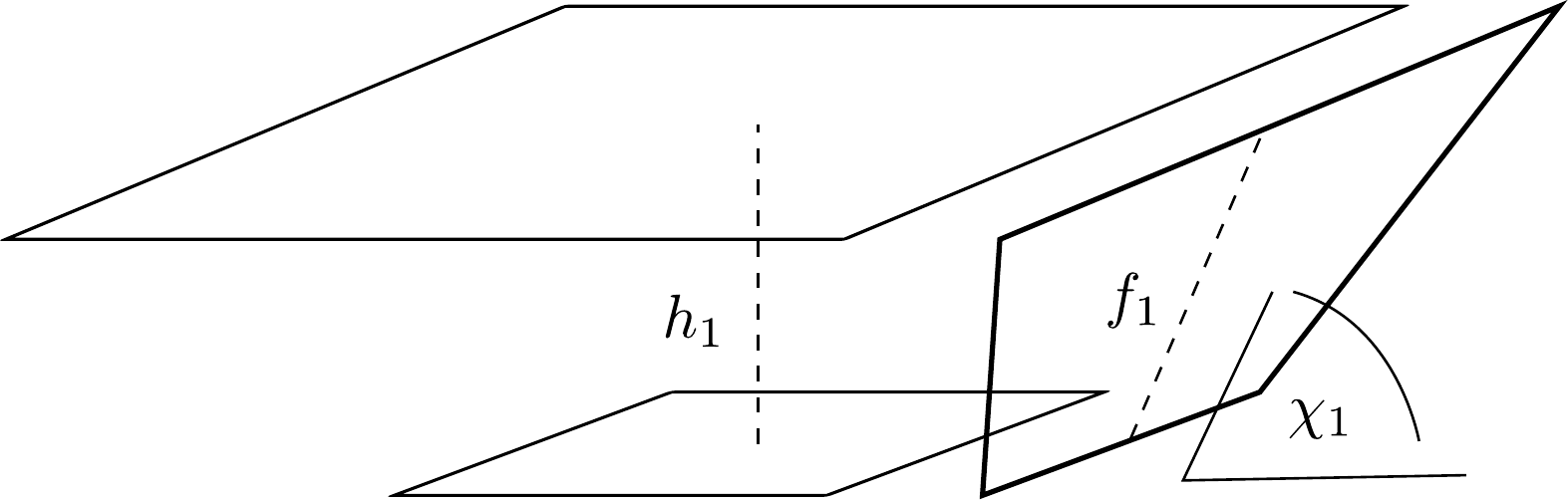}
\caption{Dimensionally reduced depiction of definition of $h_1$, which corresponds to the height  of the $4d$ anvil. The two parallel rectangles are the rotated cuboids in $\mathbb{R}^4$, the slanted trapezoid is the rotated anvil $1$. The heights $h_2$ and $h_2$ are defined analogously.}\label{Fig:4dHeight}
\end{center}
\end{figure}

\noindent The interpretation for this is that, in a $4d$ anvil the $4d$ height $h$ can be defined by any of the $3d$ heights and the $3d$ dihedral angle (see figure \ref{Fig:4dHeight}). The conditions (\ref{Eq:GeometricityConstraintAlternative}) then demand that all three ways of computing the height are equivalent. 

It is straightforward to show that the variables (\ref{Eq:GICS_Data}) subject to the constraints (\ref{Eq:EquationsAndConstraints}) and (\ref{Eq:GeometricityConstraintAlternative}) can be used to define a geometrical anvil. For this, reconstruct the metric data for each individual $3d$ anvil, and then note that the cuboids with side lengths $l_{23}$, $l_{31}$, $l_{12}$, as well as $k_{23}$, $k_{31}$, $k_{12}$, separated by a height $h$ equal to (\ref{Eq:GeometricityConstraintAlternative}) form precisely the $4d$ anvil resulting in the original spin foam data.

\subsection{Geometricity constraints for generic geometries}\label{GeoConGG}

In the case of more general geometries on the hypercuboidal graph $\Gamma$, it is unclear at this point how to formulate the geometricity constraints. However, one can estimate their number, and it coincides with the number of Hopf links in $\Gamma$ minus one. Meaning that one can hope that the general constraints can be formulated in terms of equality of quantities associated to Hopf links. These will not be the Hopf link volumes, but might be (as in the case of the anvil \eqref{Eq:HopfLinkVolumeSimplicityCheck}) expressed as slight deviations from the Hopf link volumes.

Assume there is a convex $4d$ polytope, of which the boundary graph is given by the hypercuboidal graph $\Gamma_H$. By Minkowski's theorem, this polytope -- up to translations -- is uniquely defined by the $4d$ normals $N_a$  and $3d$ volumes $v_a$ of its $3d$ boundary polytopes. Since $\Gamma_H$ has eight nodes, this amounts to 32 real variables. Due to closure, these satisfy the 4 constraints
\begin{eqnarray}
\sum_av_aN_a\;=\;0,
\end{eqnarray}

\noindent and by dividing out the 6-dimensional group of rotations in $\mathbb{R}^4$, we arrive at
\begin{eqnarray}
\dim P_{8}\;=\;22,
\end{eqnarray}

\noindent where $P_{8}$ is the space of $4d$ polytopes with a boundary graph having eight nodes.

On the other hand, we can consider the dimension of bivector geometries $\mathcal{B}_{\Gamma_H}$ on $\Gamma_H$. First we compute the dimension of gauge-invariant spin network geometries $\mathcal{S}_{\Gamma_H}$, by considering the areas $j_{ab}$ and normals $\vec{n}_{ab}$, satisfying $3d$ closure
\begin{eqnarray}
\sum_{b\supset a}j_{ab}\vec{n}_{ab}\;=\;0.
\end{eqnarray} 

\noindent With 8 nodes in the graph, and each node being six-valent, the dimension of classical spinfoam data, i.e.~the set of possibly degenerate $3d$ polytopes with matching areas, is given by
\begin{eqnarray*}
\dim \mathcal{S}_{\Gamma_H}\;&=&\;\underbrace{24}_{\#\text{ of }j_{ab}}\;+\;2\cdot \underbrace{8\cdot 6}_{\#\text{ of }\vec{n}_{ab}}\\[5pt]
&&\,-\,\underbrace{8\cdot 3}_{\text{closure}}\;-\;\underbrace{8\cdot 3}_{\text{g.i.}}\;=\;72
\end{eqnarray*}

\noindent To compute the dimension of  $\mathcal{B}_{\Gamma_H}$, we note that the bivector data is given by $\{j_{ab},\vec{n}_{ab},g_a^{\pm}\}$, with $\{j_{ab},\vec{n}_{ab}\}$ being spin foam data, and $(g_a^+,g_a^-)\in \text{Spin}(4)\simeq SU(2)\times SU(2)$ satisfying the critical and stationary equations (\ref{Eq:CritStat_SU2}), which are 4 equations per edge. We therefore have
\begin{eqnarray*}
\dim \mathcal{B}_{\Gamma_H}\;=\;\dim \mathcal{S}_{\Gamma_H}\,+\,\underbrace{8\cdot 6}_{\#\text{of }g_a^\pm}\,-\,\underbrace{24\cdot 4}_{\#\text{ crit stat}}\;=\;24
\end{eqnarray*} 

\noindent Note that the global gauge symmetry has already been taken into account by dividing out the local rotation symmetry. As one can see, one generically has
\begin{eqnarray}
\dim \mathcal{B}_{\Gamma_H}\;-\;\dim P_{8}\;=\;2,
\end{eqnarray}

\noindent so one can expect that there are two constraints which restrict the space of bivector geometries on $\Gamma_H$ to polytopes with that graph. One can easily repeat this type of calculation for arbitrary graphs $\Gamma$ with $N$ nodes and $L$ links and arrive at
\begin{eqnarray}\label{Eq:NumberGeometricityConstraints_GeneralPolytope}
\begin{aligned}
\dim \mathcal{B}_{\Gamma}\;-\;\dim P_{N}\;&=\;L-(4N-10)\\[5pt]
\;&=\;10+L-4N.
\end{aligned}
\end{eqnarray}

\noindent  However, we have made a crucial assumption here, which is that the graph $\Gamma$ is stable, i.e.~for a polytope $P$ which has $\Gamma$ as boundary graph, the graph does not change under an infinitesimal change of the 3-volumes and $4d$ normals of $P$. This is not the case for every polytope, hence not for every graph. An example for a non-stable graph is given in appendix \ref{App:StableUnstableGraphs}. At this point, it seems reasonable to assume that a graph is stable if and only if it is the boundary graph of a simple polytope, i.e.~if at every vertex of the polytope meet exactly four edges. This is for instance the case for the 4-simplex and the hypercuboid, for which one can see that (\ref{Eq:NumberGeometricityConstraints_GeneralPolytope}) gives the correct answer of 0 and 2, respectively. We hope to come back to this point in the future.

\section{Hopf link volumes \& geometricity constraints}\label{Sec:HLVolumes}

\subsection{Hopf link volume simplicity constraints}

We compare the geometricity constraints (\ref{Eq:GeometricityConstraintAlternative}) with the HLVS constraints which have been proposed in \cite{Bahr:2017ajs}. 

By Hopf link $H$ we mean a pair of loops in the graph $\Gamma$ which have linking number $\pm 1$.  
The Hopf link volume $V_{H}$ is given by
\begin{eqnarray}
V_H\;=\;\frac{1}{6}\sum_{C\between H}\sigma(C)*\left(B_1\wedge B_2\right),
\end{eqnarray}

\noindent where the sum runs over all crossings $C$ between two links in $H$, $\sigma(C)$ is the sign of the crossing, $B_{1,2}$ the two bivectors associated to the two crossing links, and $*:\wedge^4\mathbb{R}^4\to\mathbb{R}$ the Hodge operator. 

To reconstruct the bivectors
\begin{eqnarray}
B_{ab}\;\simeq\; j_{ab}(\tilde{n}^+_{ab},\,\tilde{n}_{ab}^-)
\;=\;j_{ab}\left(R_{a}^+\vec{n}_{ab},\,R_{a}^-\vec{n}_{ab}\right)
\end{eqnarray}
\noindent from the spin foam data (\ref{Eq:GICS_Data}) satisfying (\ref{Eq:CSRelationsCosines}), we note that $R_{a}^\pm$ correspond to the two different solutions for the $\chi_i$ in (\ref{Eq:CSEquations}). We choose $R^+_{a}$ to be the ones with all $\chi_i\geq 0$.\footnote{This choice effectively corresponds to a choice of orientation in $4d$, and hence to the sign of the (Hopf link) volumes. Our choice is such that these are all non-negative.}

\begin{figure}[hbt!]
\begin{center}
\includegraphics[scale=0.65]{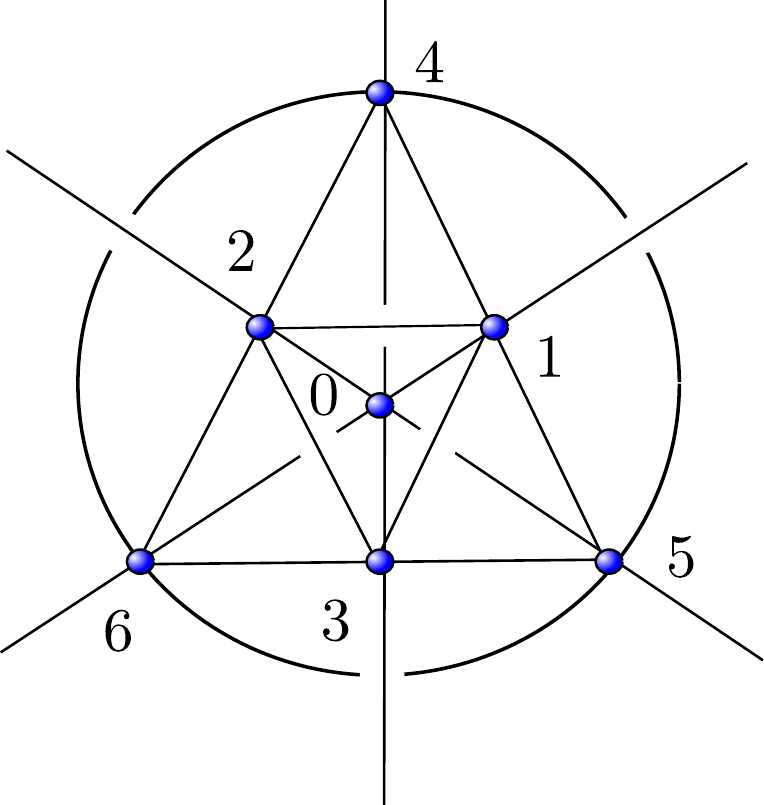}
\caption{Rearrangement of the graph from figure \ref{Fig:Anvil_4d_Graph}. Node $7$ has been moved to infinity. One can easily see the Hopf links, e.g.~$H_2$, which goes through nodes $1\to 4\to 6\to 3\to 1$ and $0\to 2\to 7\to 5\to 0$. }\label{Fig:Anvil_4d_Graph_Symmetric}
\end{center}
\end{figure}

In this graph there are three Hopf links $H_i$, $i=1,2,3$. Using (\ref{Eq:BivectorConvention01}) from the appendix, as well as figure \ref{Fig:Anvil_4d_Graph_Symmetric}, one can see that, for e.g.~for one of the crossings in Hopf link 2 one has
\begin{alignat}{2}
*\left(B_{27}\wedge B_{46}\right)&=
\frac{j_{27}j_{46}}{12} && \left(R_{2}^+\vec{n}_{27}\cdot R^+_{4}\vec{n}_{46} - R_{2}^-\vec{n}_{27}\cdot R^-_{4}\vec{n}_{46}\right) \notag \\[5pt]
%
&=
\frac{j_{27}j_{46}}{12} && \left(-\vec{n}_{72}\cdot R^+_{4}\vec{n}_{46}+\vec{n}_{72}\cdot R^-_{4}\vec{n}_{46}\right)\ \notag \\[5pt]
&=\frac{j_{27}j_{46}}{6} && \,\sin\chi_3\,\cos\phi_3,
\end{alignat}
\noindent and with (\ref{Eq:Spins}) one gets
\begin{eqnarray}
V_{H_2}\;&=&\;\frac{1}{3}j_2j_5\sin\chi_3\cos\phi_3.
\end{eqnarray}

\noindent With (\ref{Eq:Definition_X_i}) we can therefore write the Hopf link volumes as
\begin{eqnarray}\label{Eq:HopfLinkVolumes}
\begin{aligned}
V_{H_1}\;&=&\;\frac{1}{3}j_1j_6\,X_1,\\[5pt]
V_{H_2}\;&=&\;\frac{1}{3}j_2j_5\,X_2,\\[5pt]
V_{H_3}\;&=&\;\frac{1}{3}j_3j_4\,X_3.
\end{aligned}
\end{eqnarray}

\noindent A comparison with the results from \cite{Bahr:2017ajs} show that the Hopf link volumes (\ref{Eq:HopfLinkVolumes}) are generalizations of the ones for the hypercuboid, but also of those for the frusta \eqref{HLV.Frusta}.

However, the HLVS constraints $V_{H_1}=V_{H_2}=V_{H_3}$ are \emph{not} equivalent to geometricity (\ref{Eq:GeometricityConditions}). This can be seen as follows: A comparison with (\ref{Eq:PropertiesOfHeights}) yields
\begin{eqnarray}
\begin{aligned}
V_{H_1}\;&=\;\frac{j_1j_6}{z_{23}}h_1\;=\;\frac{j_1j_6}{z_{32}}h_2,\\[5pt]
V_{H_2}\;&=\;\frac{j_2j_5}{z_{13}}h_1\;=\;\frac{j_2j_5}{z_{31}}h_3,\\[5pt]
V_{H_3}\;&=\;\frac{j_3j_4}{z_{21}}h_2\;=\;\frac{j_3j_4}{z_{12}}h_1.
\end{aligned}
\end{eqnarray}

\noindent Now we assume that $z_{ij}=z_{ji}$, i.e.~geometricity holds. Then we also have $k_{ij}=k_{ji}$ and $l_{ij}=l_{ji}$, since we have shape-matching of the gluing faces. We then get
\begin{eqnarray}\label{Eq:HopfLinkVolumeSimplicityCheck}
\frac{V_{H_i}}{V_{H_j}}
\;&=&\;
\frac{\alpha_i}{\alpha_j}
\end{eqnarray} 
\noindent with
\begin{eqnarray*}
\alpha_1
\;&=&\;
\frac{(k_{12}k_{13}+l_{12}l_{13})(k_{23}+l_{23})}{4}\\[5pt]
\alpha_2
\;&=&\;
\frac{(k_{21}k_{23}+l_{21}l_{23})(k_{13}+l_{13})}{4}\\[5pt]
\alpha_3
\;&=&\;
\frac{(k_{31}k_{32}+l_{31}l_{32})(k_{21}+l_{21})}{4}
\end{eqnarray*}

\noindent By using geometricity, one can see that the expression (\ref{Eq:HopfLinkVolumeSimplicityCheck}) is equal to 1 only if either
\begin{eqnarray}
l_{12}\;=\;k_{12}\qquad\text{or}\quad \frac{k_{13}}{l_{13}}\;=\;\frac{k_{23}}{l_{23}}.
\end{eqnarray}

\noindent Geometrically this condition can be understood within the dynamics of the cuboid described by the $4d$ anvil: If the side along the $3$-direction does not expand, or if the face perpendicular to it is scaled up homogeneously. Since this is not generically the case for anvils, we conclude that the HLVS constraints are not generically equivalent with the geometricity constraints. On the other hand, in the case of hypercuboids \eqref{HLV.Hypercube} and frusta \eqref{HLV.Frusta} the conditions are satisfied, which explains why in those cases geometricity coincide with the HLVS constraints.

\subsection{The 4d anvil as a perturbation of regular configurations}\label{Sec:Perturbations}

Having shown the inequivalence between imposing the geometricity conditions and imposing the HLVS constraints in the case of the $4d$ anvil, it would be interesting to evaluate the discrepancy between the two sets of conditions. In order to do so, we adopt a perturbative approach around certain regular configurations, where the geometricity conditions and the HLVS constraints turn out to be equivalent. Namely, we analize two cases: the perturbation of the hypercuboid and the perturbation of the frusta \cite{Bahr:2017ajs}.

\subsubsection{From the hypercuboid to the 4d anvil}

We consider a perturbation of the hypercuboid, up to the second order, by perturbing the areas of the faces of the cuboids ($j_1^c,\dots,j_6^c$), and the angles $(\phi_i^c,\theta_i^c)$ which vanish for cuboids in the hypercuboid. Namely, we take
\begin{eqnarray}\label{pert.H}
\begin{aligned}
 j_i &= j_i^c + \sigma_{1,i} t + \sigma_{2,i} t^2 \\
 \phi_i &= \epsilon_{1,i} t + \epsilon_{2,i} t^2 \\
 \theta_i &= \delta_{1,i} t + \delta_{2,i} t^2
\end{aligned}
\end{eqnarray}
Our goal is to compare the geometricity conditions obtained for the 4d anvil obtained as second order perturbation of the hypercube, to the constraints which would be induced by equating Hopf link volumes. As mentioned earlier, in the case of the hypercube, the geometricity conditions are equivalent to the HLVS constraints \cite{Bahr:2017ajs}, and one obtains the conditions
\begin{align}\label{HLV.Hypercube}
 j_1^c j_6^c = j_2^c j_5^c = j_3^c j_4^c
\end{align}

Now, taking the perturbations \eqref{pert.H}, on one hand the geometricity conditions induce the following constraints
\begin{alignat}{1}
  \nonumber & 0^{th}\ \text{order:} \\ 
  & j_1^c j_6^c = j_2^c j_5^c = j_3^c j_4^c ; 
  \\
  \nonumber & \\
  \nonumber & 1^{st}\ \text{order:} \\ 
  & j_1^c \sigma_{1,6} + j_6^c \sigma_{1,1} = j_2^c \sigma_{1,5} + j_5^c \sigma_{1,2} = j_3^c \sigma_{1,4} + j_4^c \sigma_{1,3} ;
\end{alignat}
\begin{widetext}
\begingroup
\allowdisplaybreaks
\begin{alignat}{2}
  & 2^{nd}\ \text{order:} \quad && \bullet\ j_1^c j_2^c j_5^c j_6^c(\epsilon_{1,1}^2+\epsilon_{1,2}^2-\delta_{1,1}^2-\delta_{1,2}^2)+2 j_4^c j_5^c j_6^c(j_5^c \delta_{1,2} \epsilon_{1,2}-j_6^c \delta_{1,1} \epsilon_{1,1})
  \\
  \nonumber & && \ \ +2 (j_1^c \sigma_{1,6}-j_2^c \sigma_{1,5}) (j_1^c \sigma_{1,6}-j_5^c \sigma_{1,2})+2\sqrt{j_1^c j_2^c j_5^c j_6^c}(j_2^c \sigma_{2,5}+j_5^c \sigma_{2,2}-j_1^c \sigma_{2,6}-j_6^c \sigma_{2,1})=0 \ ;
  \\
  \nonumber & && \\
  & && \bullet\ j_2^c j_3^c j_4^c j_5^c(\epsilon_{1,2}^2+\epsilon_{1,3}^2-\delta_{1,2}^2-\delta_{1,3}^2)+2 j_4^c j_5^c j_6^c(j_4^c \delta_{1,3} \epsilon_{1,3}-j_5^c \delta_{1,2} \epsilon_{1,2})
  \\
  \nonumber & && \ \ +2 (j_2^c \sigma_{1,5}-j_3^c \sigma_{1,4}) (j_2^c \sigma_{1,5}-j_4^c \sigma_{1,3})+2\sqrt{j_2^c j_3^c j_4^c j_5^c}(j_3^c \sigma_{2,4}+j_4^c \sigma_{2,3}-j_2^c \sigma_{2,5}-j_5^c \sigma_{2,2})=0 \ ;
  \\
  \nonumber & && \\
  & && \bullet\ j_1^c j_3^c j_4^c j_6^c(\epsilon_{1,3}^2+\epsilon_{1,1}^2-\delta_{1,3}^2-\delta_{1,1}^2)+2 j_4^c j_5^c j_6^c(j_6^c \delta_{1,1} \epsilon_{1,1}-j_4^c \delta_{1,3} \epsilon_{1,3})
  \\
  \nonumber & && \ \ +2 (j_3^c \sigma_{1,4}-j_1^c \sigma_{1,6}) (j_3^c \sigma_{1,4}-j_6^c \sigma_{1,1})+2\sqrt{j_1^c j_3^c j_4^c j_6^c}(j_1^c \sigma_{2,6}+j_6^c \sigma_{2,1}-j_3^c \sigma_{2,4}-j_4^c \sigma_{2,3})=0 \ ;
\end{alignat}
\endgroup
\end{widetext}
where the equations for the first and second order were obtained using the previous orders equations to simplify the form of the constraints. One can further simplify the second order equations using the critical and stationary point equations \eqref{Eq:EquationsAndConstraints} which read
\begin{alignat}{1}
  & 1^{st}\ \text{order:} \nonumber \\
  & \epsilon_{1,1} = -\delta_{1,3} \ ;\ \epsilon_{1,2} = -\delta_{1,1} \ ;\ \epsilon_{1,3} = -\delta_{1,2} \label{1CSeqs}
  \\
  \nonumber & \\
  & 2^{nd}\ \text{order:}\nonumber \\
  & \epsilon_{2,1} = -\delta_{2,3} \ ;\ \epsilon_{2,2} = -\delta_{2,1} \ ;\ \epsilon_{2,3} = -\delta_{2,2}
\end{alignat}
we then get as geometricity conditions
\begingroup
\allowdisplaybreaks
\begin{alignat}{1}
  \nonumber & 0^{th}\ \text{order:} \\ 
  & j_1^c j_6^c = j_2^c j_5^c = j_3^c j_4^c ; 
  \\
  \nonumber & \\
  \nonumber & 1^{st}\ \text{order:} \\ 
  & j_1^c \sigma_{1,6} + j_6^c \sigma_{1,1} = j_2^c \sigma_{1,5} + j_5^c \sigma_{1,2} = j_3^c \sigma_{1,4} + j_4^c \sigma_{1,3} ;
  \\
  \nonumber & \\
  \nonumber & 2^{nd}\ \text{order:} \\ 
  \nonumber & \bullet\ \frac{j_1^c j_2^c j_5^c j_6^c}{2} (\epsilon_{1,1}^2-\epsilon_{1,3}^2)+ j_4^c j_5^c j_6^c \epsilon_{1,2}(j_6^c \epsilon_{1,1} - j_5^c \epsilon_{1,3} ) 
  \\
  \nonumber & \ \ \ + \sqrt{j_1^c j_2^c j_5^c j_6^c}(j_2^c \sigma_{2,5}+j_5^c \sigma_{2,2}-j_1^c \sigma_{2,6}-j_6^c \sigma_{2,1})
  \\ 
  & \ \ \ + (j_1^c \sigma_{1,6}-j_2^c \sigma_{1,5}) (j_1^c \sigma_{1,6} - j_5^c \sigma_{1,2}) = 0\ ;
  \\
  \nonumber & \\
  \nonumber & \bullet\ \frac{j_2^c j_3^c j_4^c j_5^c}{2} (\epsilon_{1,2}^2-\epsilon_{1,1}^2)+ j_4^c j_5^c j_6^c \epsilon_{1,3}(j_5^c \epsilon_{1,2} - j_4^c \epsilon_{1,1}) 
  \\
  \nonumber & \ \ \ + \sqrt{j_2^c j_3^c j_4^c j_5^c}(j_3^c \sigma_{2,4}+j_4^c \sigma_{2,3}-j_2^c \sigma_{2,5}-j_5^c \sigma_{2,2})
  \\
  & \ \ \ + (j_2^c \sigma_{1,5}-j_3^c \sigma_{1,4}) (j_2^c \sigma_{1,5}-j_4^c \sigma_{1,3}) = 0 \ ;
  \\
  \nonumber & \\
  \nonumber & \bullet\ \frac{j_1^c j_3^c j_4^c j_6^c}{2} (\epsilon_{1,3}^2-\epsilon_{1,2}^2)+ j_4^c j_5^c j_6^c \epsilon_{1,1} (j_4^c \epsilon_{1,3} - j_6^c \epsilon_{1,2}) 
  \\
  \nonumber & \ \ \ + \sqrt{j_1^c j_3^c j_4^c j_6^c}(j_1^c \sigma_{2,6}+j_6^c \sigma_{2,1}-j_3^c \sigma_{2,4}-j_4^c \sigma_{2,3})
  \\
  & \ \ \ + (j_3^c \sigma_{1,4}-j_1^c \sigma_{1,6}) (j_3^c \sigma_{1,4}-j_6^c \sigma_{1,1}) = 0 \ ;
\end{alignat}
\endgroup
where now it is clear that the three second order conditions are not independent when assuming \eqref{1CSeqs}, and only two of them are.

On the other hand, equating the Hopf link volumes give the following constraints
\begingroup
\allowdisplaybreaks
\begin{alignat}{1}
  \nonumber & 0^{th}\ \text{order:}\\ & j_1^c j_6^c = j_2^c j_5^c = j_3^c j_4^c;
  \\
  \nonumber & \\
  \nonumber & 1^{st}\ \text{order:}\\ & j_1^c \sigma_{1,6} + j_6^c \sigma_{1,1} = j_2^c \sigma_{1,5} + j_5^c \sigma_{1,2} = j_3^c \sigma_{1,4} + j_4^c \sigma_{1,3};
  \\
  \nonumber & \\
  \nonumber & 2^{nd}\ \text{order:} \\ 
  & j_1^c j_6^c \delta_{1,3}^2 + j_1^c j_6^c \epsilon_{1,2}^2 - 2( \sigma_{1,1} \sigma_{1,6}+j_6^c \sigma_{2,1}+j_1^c \sigma_{2,6})
  \\
  \nonumber & = j_2^c j_5^c \delta_{1,1}^2 + j_2^c j_5^c \epsilon_{1,3}^2 - 2(\sigma_{1,2} \sigma_{1,5}+j_5^c \sigma_{2,2}+j_2^c \sigma_{2,5})
  \\
  \nonumber & = j_3^c j_4^c \delta_{1,2}^2 + j_3^c j_4^c \epsilon_{1,1}^2 - 2(\sigma_{1,3} \sigma_{1,4}+j_4^c \sigma_{2,3}+j_3^c \sigma_{2,4});
\end{alignat}
\endgroup
One can then observe that while the HLVS constraints induce the geometricity conditions at the zeroth order (expected) and first order, they do not generate the geometricity conditions at a higher order, examplified by the second order equations above. Leading to the conclusion that the HLVS constraints can be viewed as a first order approximation of the geometricity conditions in the vicinity of the hypercube configuration.

\subsubsection{From the frusta to the 4d anvil}

This time we consider perturbations of the areas of the faces of the frusta ($j_1^c=j_2^c=j_3^c=j^c,j_4^c=j_5^c=j_6^c=j^s$) and its angles $(\phi_i^c=-\theta_i^c=\zeta)$, which implies that the $4d$ angles are all equal $\chi_i=\chi$.  Similarly to \eqref{pert.H}, but only up to first order as it is sufficient for the analysis, we take
\begin{eqnarray}\label{pert.F}
\begin{aligned}
 j_i &= j^c + \sigma_{1,i} t\ ,\qquad i\in \{1,2,3\} \\
 j_k &= j^s + \sigma_{1,k} t\ ,\qquad k\in \{4,5,6\} \\
 \phi_i &= \zeta + \epsilon_{1,i} t\ ,\ \theta_i = -\zeta + \delta_{1,i} t
\end{aligned}
\end{eqnarray}
In the case of the frusta, the geometricity conditions are trivially satisfied after imposing the critical and stationary point conditions \eqref{Eq:EquationsAndConstraints}, and the Hopf link volumes coincide, namely
\begin{align}\label{HLV.Frusta}
 V_{H_i}=j^c j^s \sin\chi \sin\zeta
\end{align}
Introducing the perturbations \eqref{pert.F}, the geometricity conditions become constraints on the perturbation parameters $(\sigma,\epsilon,\delta)$, for instance the equation $z_{12}=z_{21}$ induces
\begin{widetext}
\begin{alignat}{1}\label{Geom.Cond.Frusta}
  & \bigg(j^c - \sqrt{(j^c)^2-4 (j^s \sin(\zeta))^2} \bigg) \bigg(2 j^c (\delta_{1,1}-\epsilon_{1,2})+2 j^c (\delta_{1,2}-\epsilon_{1,1}) \cos(2 \zeta)+(\sigma_{1,1}+\sigma_{1,2}) \sin(2 \zeta) \bigg) \\ \nonumber 
  &=
  2 j^s \sin(\zeta)^2 \bigg(j^s (3 \delta_{1,1}-\delta_{1,2}+\epsilon_{1,1}-3 \epsilon_{1,2})+j^s \cos(2\zeta) (-\delta_{1,1}+3 \delta_{1,2}-3 \epsilon_{1,1}+\epsilon_{1,2}) + \sin(2\zeta) (2 \sigma_{1,4}+\sigma_{1,5}+\sigma_{1,6}) \bigg)\ ;
\end{alignat}
\end{widetext}
The other two equations are obtained by a simple cyclic permutation of the indices of the perturbation parameters.
On one hand, the critical and stationary point equations \eqref{Eq:EquationsAndConstraints} in this case read
\begin{alignat}{1}
  & (\delta_{1,2}+\epsilon_{1,3}) \cos^2(\zeta) = (\delta_{1,1}+\epsilon_{1,1}) \sin^2(\zeta)
  \\
  & (\delta_{1,3}+\epsilon_{1,1}) \cos^2(\zeta) = (\delta_{1,2}+\epsilon_{1,2}) \sin^2(\zeta)
  \\
  & (\delta_{1,1}+\epsilon_{1,2}) \cos^2(\zeta) = (\delta_{1,3}+\epsilon_{1,3}) \sin^2(\zeta)
\end{alignat}
which implies
\begin{align}
 \epsilon_{1,1}+\epsilon_{1,2}+\epsilon_{1,3}= - (\delta_{1,1}+\delta_{1,2}+\delta_{1,3})
\end{align}
On the other hand, the Hopf link volumes constraints reduce to
\begin{align}\label{HLVC.Frusta}
  \nonumber & j^c j^s (\delta_{1,3}-\epsilon_{1,2}) \sin (2 \zeta )+2 \cos (2 \zeta ) (j^c \sigma_{1,6}+j^s \sigma_{1,1})
  \\
  \nonumber & = j^c j^s (\delta_{1,1}-\epsilon_{1,3}) \sin (2 \zeta )+2 \cos (2 \zeta ) (j^c \sigma_{1,5}+j^s \sigma_{1,2})
  \\
  & = j^c j^s (\delta_{1,2}-\epsilon_{1,1}) \sin (2 \zeta )+2 \cos (2 \zeta ) (j^c \sigma_{1,4}+j^s \sigma_{1,3})
\end{align}
The comparison of \eqref{Geom.Cond.Frusta} and \eqref{HLVC.Frusta} shows that the HLVS constraints for the frusta do not generate the geometricity conditions, not even at first order, unlike the hypercube case. Hence we establish that in the case of the frusta, the HLVS constraints cannot be viewed as an approximation of the geometricity conditions in the vicinity of the frusta configuration.

\section{Summary and outlook}\label{Sec:Summary}

In this article, we analyzed the semi-classical limit of the transition amplitude corresponding to a four dimensional anvil in the EPRL-KKL spin foam model. Using the bivector geometry associated to the $4d$ anvil boundary graph, we particularly explored the geometricity conditions required to recover the anvil geometry, and their relation to the HLVS constraints.

We started by formulating the geometry of the boundary of a $4d$ anvil both in terms of the data of a spin network, i.e.\ the spins and normals associated to the boundary graph of the anvil and satisfying the closure conditions, and in terms of the metric variables which consist of edge lengths and heights. We then defined the bivector geometry associated to this boundary graph by imposing the critical and stationary point conditions. Next we looked into the geometricity conditions which stand for the constraints reducing a bivector geometry to a $4d$ polytope. We found that in the case of the anvil we recover two independent conditions, which simply state that the four dimensional height of the $4d$ anvil can be computed from any 3d anvil data on the boundary, and the different expressions would give the same value. After that, we moved to the calculation of the Hopf link volumes associated to the $4d$ anvil boundary graph, and we compared the geometricity conditions to the HLVS constraints, which consist of equating the Hopf link volumes. We consequently established that these HLVS constraints are not equivalent to geometricity, unlike what happens in the case of the hypercube \cite{Bahr:2017ajs}, though we manage to reexpress the the geometricity conditions in terms of the Hopf link volumes. We further investigate the discrepancy between these two sets of conditions via a perturbative analysis, where the $4d$ anvil geometry is considered to be obtained as a perturbation of a hypercube, and of a frusta. The perturbative analysis yields the conclusion that in the vicinity of the hypercube configuration, the HLVS constraints reproduce geometricity up to the first order in the perturbation parameter, while the approximation fails in the vicinity of the frusta, showing that from this perspective the frusta is rather an unstable configuration.

In conclusion, we have shown that the geometricity conditions for a given boundary graph do not coincide in general with the HLVS constraints. However, it remains that in terms of the counting of degrees of freedom, the HLVS constraints associated to the boundary graph of $4d$ anvil expose the correct number of non geometrical degrees of freedom present in the bivector geometry. Taking into account the hypercube and frusta cases, and adding to that the fact that one can express the geometricity conditions for the 4d anvil in terms of Hopf link volumes, the analysis seem to suggest that the Hopf link volumes are strongly tied to the geometricity of a bivector geometry. This connection needs to be further investigated in the cases of more general configurations, i.e.\ transition amplitudes with more complicated boundary graphs. Finally, as suggested by the discussion about the counting of degrees of freedom in section \ref{GeoConGG}, determining and controlling the geometricity conditions for an arbitrary boundary graph will require a more complex analysis, where for instance we might expect the degenerate configurations to be treated on the same footing as the non degenerate ones. We hope to investigate these questions in detail in future works.

\acknowledgments

This work was funded in part by the project BA 4966/1-1 of the German Research Foundation (DFG), and by the OPUS 15 Grant nr 2018/29/B/ST2/01250 of the Polish National Science Center. The authors also acknowledge the support of the DFG under Germany's Excellence strategy EXC2121 (Quantum Universe) 390833306.

\appendix 

\section{Bivector conventions}\label{App:SelfDualConventions}

A bivector $B_{ab}=-B_{ba}\in\bigwedge^2\mathbb{R}^4$, $a,b=0,1,2,3$, can be dualized via the Hodge operator
\begin{eqnarray}
(*B)_{ab}\;:=\;\frac{1}{2}\epsilon_{abcd}B_{cd},
\end{eqnarray}

\noindent where indices are raised and lowered with the Kronecker delta $\delta_{ab}$. The Killing form on $\bigwedge^2\mathbb{R}^4$ is taken to be positive definite as
\begin{eqnarray}
\langle B_1,B_2\rangle\;:=\;-\frac{1}{4}\text{tr}(B_1B_2).
\end{eqnarray}

\noindent The isomorphism $\bigwedge^2\mathbb{R}^4\simeq \mathbb{R}^3\oplus\mathbb{R}^3$ 
\begin{eqnarray}
B\;\leftrightarrow\;(\vec{b}^+,\,\vec{b}^-)
\end{eqnarray}

\noindent is given by 
\begin{eqnarray}
b^{\pm,I}\;=\;\frac{1}{2}\left(B_{0I}\pm\frac{1}{2}\epsilon_{IJK}B_{JK}\right)
\end{eqnarray}

\noindent with $I=1,2,3$. The wedge product of two bivectors $B$ and $C$ is defined to be
\begin{eqnarray}
(B\wedge C)_{abcd}\;=\;\frac{1}{24}\epsilon_{abcd}\epsilon^{efgh}B_{ef}C_{gh}.
\end{eqnarray}

\noindent Acting with the Hodge dual on this yields a number which is
\begin{eqnarray}\label{Eq:BivectorConvention01}
\begin{aligned}
*(B\wedge C)\;&=\;\frac{1}{24}\epsilon^{efgh}B_{ef}C_{gh}\\[5pt]
&=\;\frac{1}{12}\left(\vec{b}^+\cdot \vec{c}^+-\vec{b}^-\cdot \vec{c}^-\right),
\end{aligned}
\end{eqnarray}

\noindent and which can be regarded as the expression for the 4d volume in the volume simplicity constraint \cite{Engle:2007wy, Bahr:2017ajs}.

\section{Stable and unstable polytopes}\label{App:StableUnstableGraphs}

A polytope $P$ is, by Minkowski's theorem, given by 3-volumes $V_a$ of its 3-faces, as well as their $4d$ normals $N_a$, subject to closure $\sum_a V_a N_a=0$. We call a polytope stable if its boundary graph does not change under any infinitesimal change of the data $V_a, N_a$. We call a graph stable if it is the boundary graph of a stable polytope.

Examples for stable polytopes are the 4-simplex, or the hypercuboid.

\begin{figure}[hbt!]
\begin{center}
\includegraphics[scale=0.55]{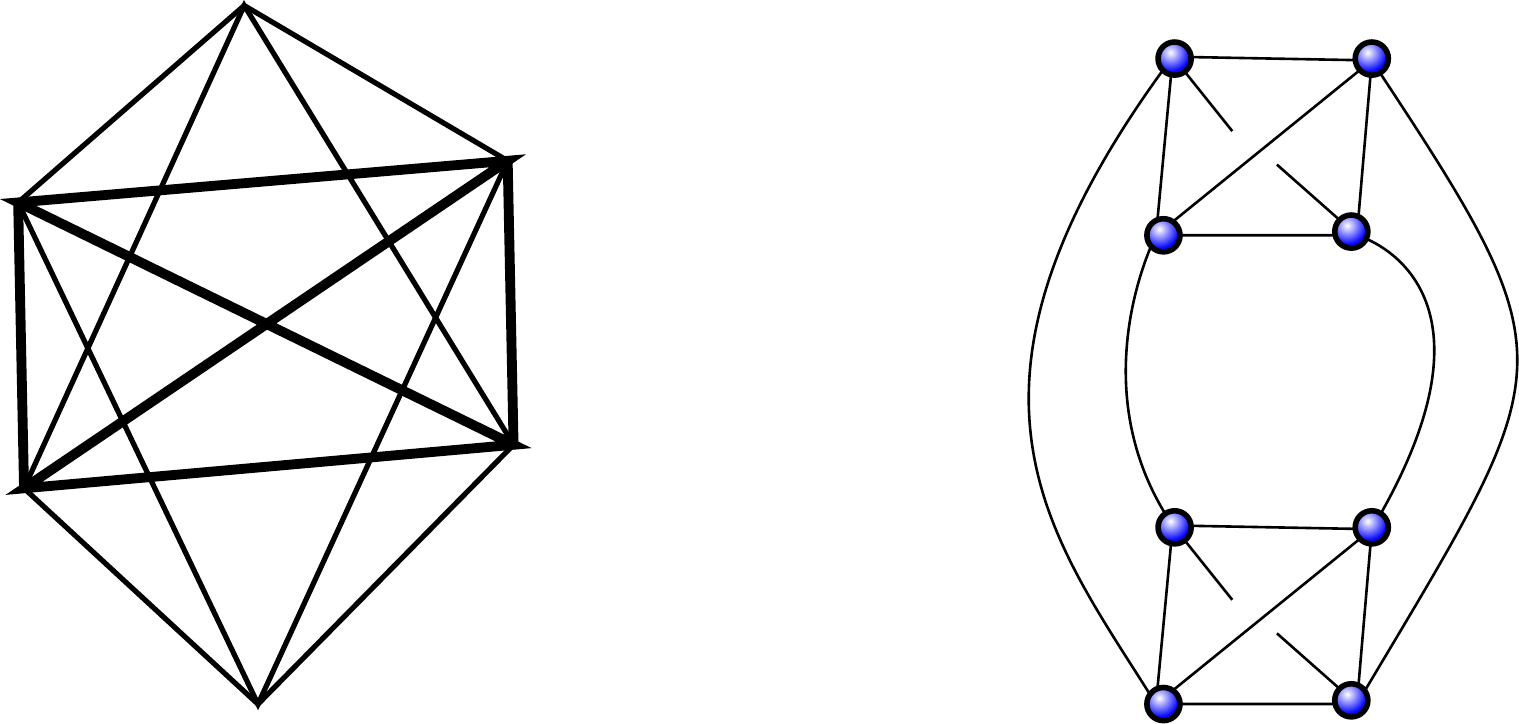}
\caption{Example for a non-simple polytope (left), obtained by gluing two 4-simplices together along a common tetrahedron (bold lines). Its boundary graph (right) is not stable. }\label{Fig:Double_Simplex}
\end{center}
\end{figure}

In the following we will give an example of a polytope which is not stable: Consider the $4d$ polytope obtained by gluing two 4-simplices along a common tetrahedron (see figure \ref{Fig:Double_Simplex}). Its boundary graph is $\Gamma_{2S}$. This graph has eight nodes, and from Minkowski's theorem we know that the space of $4d$ polytopes with eight nodes $\mathcal{P}_8$ is 22-dimensional. However, it is evident that not every prescription of $4d$ normals and $3d$ volumes to the nodes of $\Gamma_{2S}$ would give the $4d$ polytope obtained by gluing two 4-simplices. This can be seen as follows: Given data $N_a, V_a$, $a=1,\ldots,8$ on $\Gamma_{2S}$ one can construct the data of two 4-simplices by specifying them by
\begin{eqnarray*}
V_1N_1,\,V_2N_2,\,V_3 N_3,\,V_4N_4,\,-\sum_{i=1}^4V_iN_i,\\[5pt]
V_5N_5,\,V_6N_6,\,V_7 N_7,\,V_8N_8,\,-\sum_{i=5}^8V_iN_i
\end{eqnarray*}

\noindent respectively. This gives that data of two 4-simplices, which have, under gluing them together on the common tetrahedron, to result in the original double-simplex. However, nothing guarantees that the two $4$-simplices can actually be glued together, since the two tetrahedra can have entirely different shapes, even different face areas. This means not every set of data can give a double-simplex, and hence have $\Gamma_{2S}$ as boundary graph. Indeed, up to translations and rotations, the set of all possible polytopes of that type is only 14-dimensional\footnote{This can be seen easily by noting that one 4-simplex is determined by 10 edge lengths, plus the position of one additional vertex in $\mathbb{R}^4$, or, equivalently, the data of two 4-simplices ($2\times 10$), minus the six conditions that the gluing tetrahedrons agree.}.

This means that for a double-simplex, there are always infinitesimal changes of the boundary data $V_a, N_a$ such that the boundary graph is changed. This is a phenomenon which occurs for certain polytopes in any dimension greater than two, so we give an example in $d=3$. Any higher dimension works analogously.

\begin{figure}[hbt!]
\begin{center}
\includegraphics[scale=0.35]{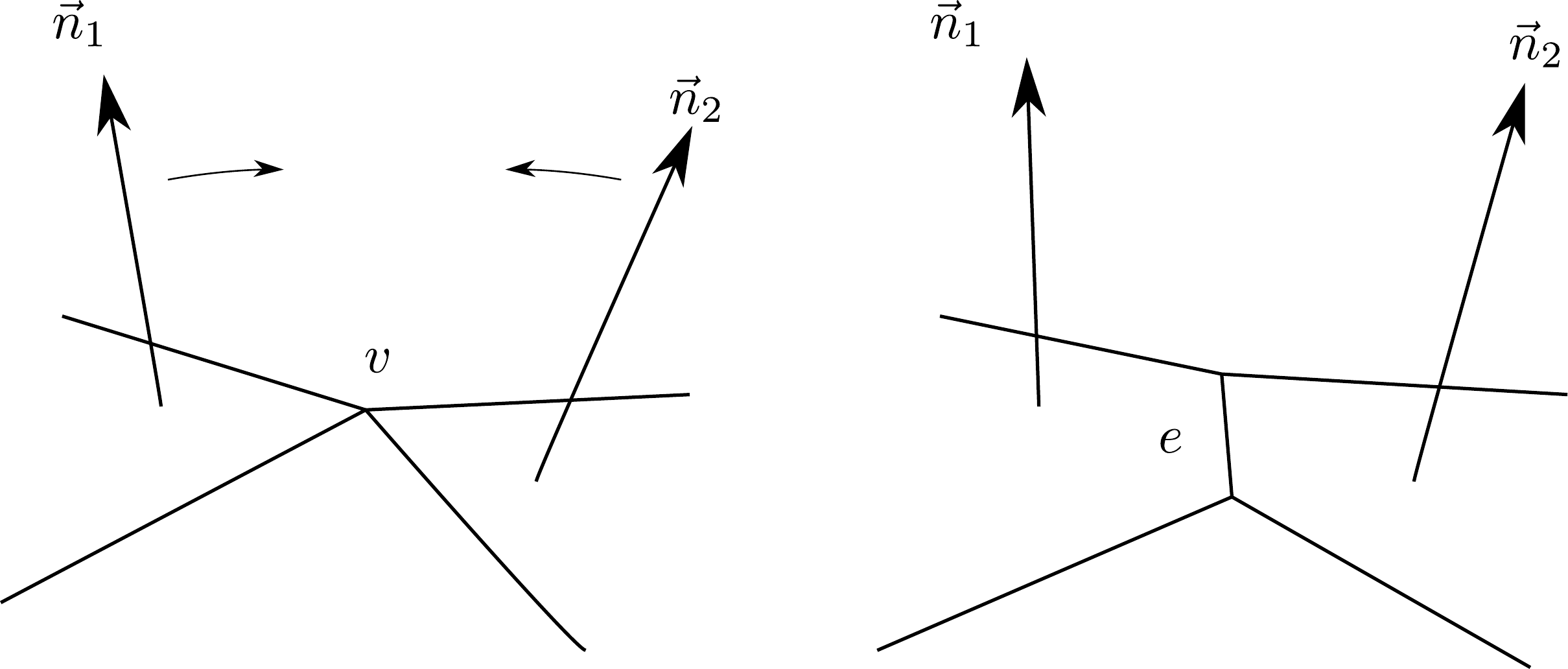}
\caption{Example for a non-simple polytope in $3d$, whose boundary graph changes under even a slight deformation of the face normals, which generates a new edge $e$, hence an new link in the boundary graph.}\label{Fig:Not_Simple_Polytope_3d}
\end{center}
\end{figure}

Consider a $3d$ polytope which is not simple, i.e.~there is at least one vertex $v$ on the boundary which has more than three edges $e_i$ attached to it (see figure \ref{Fig:Not_Simple_Polytope_3d}). Then in the boundary graph (which is now embedded in $S^2$), each face surrounding the vertex corresponds to a node, and each of these nodes is connected to its nearest neighbor, but none other of these. Now, on the side of the polytope, consider a deformation of the face normals and face areas, such that for normals associated to two non-adjacent nodes $n_1$ and $n_2$ (which have to exist, because the polytope is not simple), are rotated towards one another. As a result, even under the slightest rotation, the vertex $v$ splits up into two vertices, with an additional edge $e$ between them. On the side of the boundary graph, $e$ corresponds to an additional link, which forms between $n_1$ and $n_2$. Thus, even the slightest deformation of polytope data changes the graph.\\[5pt]

One can see that this effect cannot occur when the polytope is simple, i.e.~every vertex has exactly three edges attached to it. Similarly, in $d$ dimensions, a polytope vertex needs to have exactly $d$ edges meeting at it, which is true for e.g.~the $4$-simplex and the hypercuboid. This suggests that the simplicity of polytopes can be translated into the stability of boundary graphs under infinitesimal deformation, and we conjecture that this is indeed the case. It would be interesting to come back to this point in the future.

\bibliography{bibliography}
\bibliographystyle{ieeetr}

\end{document}